\documentclass[onecolumn,notitlepage,nofootinbib,superscriptaddress,noeprint,aps,pra]{revtex4-2}
\pdfoutput=1
\usepackage[utf8]{inputenc}
\usepackage[english]{babel}
\usepackage[T1]{fontenc}
\usepackage{microtype}
\usepackage{amsmath,amssymb,bbm,soul}
\usepackage{caption}
\usepackage{graphicx}
\usepackage{float}
\usepackage[dvipsnames]{xcolor}
\usepackage[breaklinks=true,colorlinks=true,linkcolor=teal,urlcolor=teal,citecolor=teal]{hyperref}
\usepackage{orcidlink}

\catcode`\|=\active \def|{
\fontencoding{T1}\selectfont\symbol{124}\fontencoding{\encodingdefault}}

\usepackage[normalem]{ulem}
\newcommand{\stkout}[1]{\ifmmode\text{\sout{\ensuremath{#1}}}\else\sout{#1}\fi}

\newcommand{\ket}[1]{|#1\rangle} %ket
\newcommand{\bra}[1]{\langle#1|} %bra

\DeclareMathOperator{\Tr}{Tr}

\let\Re\relax
\let\Im\relax
\DeclareMathOperator{\Re}{Re}
\DeclareMathOperator{\Im}{Im}

\begin{document}

\title{Pseudomode treatment of strong-coupling quantum thermodynamics}

\author{Francesco Albarelli\,\orcidlink{0000-0001-5775-168X}}
\email{francesco.albarelli@gmail.com}
\affiliation{Dipartimento di Fisica ``Aldo Pontremoli'', Università degli Studi di Milano, via Celoria 16, 20133 Milan, Italy}
\affiliation{Scuola Normale Superiore, I-56126 Pisa, Italy}

\author{Bassano Vacchini\,\orcidlink{0000-0002-7574-9951}}
\email{bassano.vacchini@mi.infn.it}
\affiliation{Dipartimento di Fisica ``Aldo Pontremoli'', Università degli Studi di Milano, via Celoria 16, 20133 Milan, Italy}
\affiliation{Istituto Nazionale di Fisica Nucleare, Sezione di Milano, via Celoria 16, 20133 Milan, Italy}

\author{Andrea Smirne\,\orcidlink{0000-0003-4698-9304}}
\email{andrea.smirne@unimi.it}
\affiliation{Dipartimento di Fisica ``Aldo Pontremoli'', Università degli Studi di Milano, via Celoria 16, 20133 Milan, Italy}
\affiliation{Istituto Nazionale di Fisica Nucleare, Sezione di Milano, via Celoria 16, 20133 Milan, Italy}

\begin{abstract}
The treatment of quantum thermodynamic systems beyond weak coupling is of increasing relevance, yet extremely challenging.
The evaluation of thermodynamic quantities in strong-coupling regimes requires a nonperturbative knowledge of the bath dynamics, which in turn relies on heavy numerical simulations.
To tame these difficulties, considering thermal bosonic baths linearly coupled to the open system, we derive expressions for heat, work, and average system-bath interaction energy that only involve the autocorrelation function of the bath and two-time expectation values of system operators.
We then exploit the pseudomode approach, which replaces the physical continuous bosonic bath with a small finite number of damped, possibly interacting, modes, to numerically evaluate these relevant thermodynamic quantities.
We show in particular that this method allows for an efficient numerical evaluation of thermodynamic quantities in terms of one-time expectation values of the open system and the pseudomodes.
We apply this framework to the investigation of two paradigmatic situations.
In the first instance, we study the entropy production for a two-level system coupled to an ohmic bath, simulated via interacting pseudomodes, allowing for the presence of time-dependent driving.
Secondly, we consider a quantum thermal machine composed of a two-level system interacting with two thermal baths at different temperatures, showing that an appropriate sinusoidal modulation of the coupling with the cold bath only is enough to obtain work extraction.
\end{abstract}

\maketitle

\section{Introduction}

Thermodynamics can be regarded as one of the most successful and useful branches of physics, finding application in many different scientific and technical fields.
Although it was initially developed to address concrete technological applications, it encompasses deep foundational questions, such as the relationship between the microscopic and the macroscopic domains.
Going beyond classical physics, quantum thermodynamics seeks to combine concepts from the classical theory with a quantum mechanical description of systems and their dynamics.

In the paradigmatic description of a quantum thermal machine, the working medium is a quantum system coupled to thermal reservoirs; the theory of open quantum systems~\cite{breuer2002theory,Vacchini2024} provides the tools to deal precisely with such scenarios.
Indeed, the initial investigations about the interplay between thermodynamic concepts and quantum dynamics arose in the open quantum systems community decades ago~\cite{Alicki1979}.
However, only in the past decade quantum thermodynamics has blossomed and evolved into a fully-fledged research field~\cite{Vinjanampathy2015,Binder2018,Deffner2019,Potts2024}, branching into several directions.
On the one hand, a more abstract approach has stressed the interplay with quantum information processing, mostly distancing itself from specific models~\cite{Goold2016,Lostaglio2018}.
On the other hand, a more pragmatic approach focuses on engineering and realizing useful quantum thermodynamic devices~\cite{Myers2022a}, beyond the current generation of proof-of-concept experiments.
More in general, in the whole field of quantum technologies, which is rapidly advancing, thermodynamic and energetic considerations are starting to become operationally relevant~\cite{Auffeves2022,Arrachea2023}.

In this work, we tackle the problem of computing thermodynamic quantities from a \emph{dynamical} point of view, by considering the nonperturbative evolution of an open quantum system that interacts strongly with one or multiple thermal baths.
Since the dynamics of the system and baths is customarily modeled as purely unitary, the standard definitions of heat and work for unitary systems can be applied~\cite{Kato2016,Wiedmann2020a,Liu2021a,Landi2021a}.
In this framework heat corresponds to the variation of a bath's average energy, while work corresponds to the variation of the total (system and bath) average energy due to an explicit time-dependence in Hamiltonian, which describes the influence of an external (classical) agent who supplies or collects energy to or from the system and baths.\footnote{
In this work we will only consider \emph{average} thermodynamic quantities. 
In fact, following ideas in stochastic thermodynamics, the probability distribution of heat and work can be defined, and higher-order moments studied.
However, even for unitary dynamics, there exist nonequivalent ways to proceed beyond average values, and considerations about measurements become crucial~\cite{Perarnau-Llobet2017,Talkner2020a}.}
Crucially, to evaluate these quantities one needs to track the values of bath observables and the reduced dynamics of the open quantum system is generally not enough.
This is at odds with the weak-coupling regime, where the interaction energy is negligible and heat and work become properties of the open quantum system only.
It is to be stressed that proposals have been put forward to define thermodynamic quantities through the open system's dynamics only, even for the case of arbitrary strong coupling, see e.g. Refs.~\cite{Colla2022,Picatoste2024,Alipour2020,Alipour2022}.

In this work, we will consider bosonic baths initially in a thermal state and linearly coupled to the open quantum system: this guarantees that the dynamics of the open system is fully determined by the autocorrelation function of the bath.
Under these assumptions, we can introduce our first result, which consists in explicit expressions for heat, work and system-bath average interaction energy in terms of two-time correlation functions of the open quantum system, Eqs.~\eqref{eq:qexpc}--\eqref{eq:iexpc}.
While similar results have appeared previously~\cite{Kato2016,Gribben2022}, the present derivation may be of independent interest, in that it provides a strategy combining Heisenberg's equations of motion with the thermofield approach.

While the result mentioned above is formally exact, to concretely compute those figures of merit beyond weak coupling, the dynamics must be studied almost inevitably by means of numerical simulations.
Various techniques have been proposed for this task, by adapting and extending numerical methods originally devised to study only the reduced dynamics of the open system.
These include:
hierarchical equations of motions~\cite{Kato2016,Cerrillo2016,Koyanagi2022a,Koyanagi2022,Latune2023}, reaction-coordinate mapping~\cite{Strasberg2016a,Newman2017,Newman2020,Anto-Sztrikacs2021a,Anto-Sztrikacs2022,Anto-Sztrikacs2023}, tensor-network techniques based on the Feynman-Vernon path-integral approach~\cite{Brenes2020,Popovic2021a,Gribben2022,Chen2023e}, stochastic Liouville-von Neumann equations~\cite{Wiedmann2020a,Wiedmann2021} and hierarchy of pure states~\cite{Boettcher2024}.
We also mention that for Gaussian systems of harmonic oscillators, semi-analytical solutions are possible~\cite{Freitas2016,Aguilar2022a,Cavaliere2022}.

In this paper we focus on the pseudomode approach to simulate the dynamics of an open quantum system~\cite{Garraway1997,Tamascelli2017}.
This strategy provides a Markovian embedding, in which the non-Markovian dynamics of the open quantum system is effectively described by the Markovian dynamics of a larger system.
More specifically, as long as the autocorrelation function of the two environments is the same, the exact reduced dynamics is identical to the one obtained by replacing the physical environment with an effective \emph{dissipative} environment, i.e. the so-called pseudomodes, whose free dynamics is governed by a Lindblad master equation.

Within the pseudomode framework, we show that the quantum state of system and pseudomodes contains enough information to conveniently compute work, heat and interaction energy in terms of explicit expressions that only involve one-time expectation values of observables.
Closely related results have appeared in Refs.~\cite{Lacerda2023a,Lacerda2023}, where the focus is on fermionic systems (known as the ``mesoscopic leads'' approach) and Ref.~\cite{Menczel2024} which focuses on non-physical pseudomodes governed by non-Hermitian Hamiltonians.
In particular, a crucial result for our treatment first appeared in Ref.~\cite{Menczel2024} relying on the use of functional derivatives of time-ordered exponentials, for which we provide an alternative derivation based on Heisenberg's equation of motion and the formalism of Refs.~\cite{Tamascelli2017,Smirne2022a}.

We showcase our pseudomode approach to strong-coupling quantum thermodynamics by means of two applications.
We first study the entropy production of a spin-boson model with ohmic spectral density, by approximating the environment with interacting pseudomodes~\cite{Mascherpa2019}.
In particular, as done in Ref.~\cite{Colla2021} but for a harmonic oscillator, we compare the definition of entropy production based on the open system dynamics~\cite{Spohn1978,Deffner2011} to the global definition involving also the bath degrees of freedom~\cite{Esposito2010,Landi2021a}.
We show that indeed the two definitions match in the weak-coupling limit, also in the presence of external time-dependent driving on the system.
The second result refers to the realm of strong-coupling and non-Markovian quantum thermal machines, which has recently attracted great interest, see e.g. Refs.~\cite{Strasberg2016a,Abiuso2019,Wiedmann2020a,Latune2023,Boettcher2024} or Ref.~\cite{Bhattacharjee2021} and references therein.
In analogy to Ref.~\cite{Cavaliere2022}, we show that it is possible to obtain a thermal machine just by performing a periodic driving of the coupling term in a spin-boson model coupled to two thermal baths.
We stress that also in this case the previous results have been obtained for a harmonic oscillator coupled to Gaussian baths.

The rest of the paper is structured as follows.
In Sec.~\ref{sec:Qthermo_unitary} we revise quantum thermodynamics in the context of a global system-baths unitary dynamics, then we provide expressions of the relevant quantities in terms of two-time correlations functions.
In Sec.~\ref{sec:pseudomode}, we first revise the pseudomode framework for open quantum systems and then we provide the expressions of the same thermodynamic quantities in term of one-time expectation values.
In Sec.~\ref{sec:examples} we present the two exemplary applications described above.
In Sec.~\ref{sec:conclusions} we critically summarize our results and point to perspectives for future work.

\section{Quantum thermodynamics with unitary system-baths dynamics}
\label{sec:Qthermo_unitary}

\subsection{General model with bosonic thermal baths}
We consider a generic finite-dimensional open system $S$ coupled to $N_B$ bosonic baths (we keep this fully general for now, but in the examples we will only consider $N_B=1,2$).
Crucially, we include the possibility of time-dependent terms both in the system Hamiltonian and in the coupling Hamiltonian.
The former allows us to describe arbitrary control strategies implemented via a time-dependent driving on the open system.
The latter plays an important role when moving beyond weak coupling. 
For example, a time-dependent interaction Hamiltonian can describe finite-time thermodynamic cycles in strongly-coupled quantum thermal machines~\cite{Wiedmann2020a,Boettcher2024}, where the corresponding work contribution cannot be neglected.

The total system-baths Hamiltonian reads ($\hbar = 1$)
\begin{eqnarray}
  \widehat{H}_{\mathrm{tot}}(t) &=& \widehat{H}_S(t)
  +\sum^{N_B}_{j=1} \widehat{H}_{B,j}
  +\sum^{N_B}_{j=1} \widehat{H}_{I,j}(t) \nonumber\\
  \widehat{H}_{B,j} &=& \sum_k \omega_{j,k}\widehat{b}^{\dag}_{j,k}\widehat{b}_{j,k} \nonumber\\
  \widehat{H}_{I,j}(t)&=&\lambda_j(t) \widehat{A}_{S,j}\otimes\sum_k g_{j, k}\left(\widehat{b}_{j,k}+\widehat{b}^\dag_{j,k}\right), 
  \label{eq:mainh}
\end{eqnarray}
where $\widehat{H}_S(t)$ is the free system Hamiltonian, which may include a classical driving term, and $\widehat{H}_{B,j}$ is the free Hamiltonian of the bath $j$, where $\widehat{b}_{j,k}$ and $\widehat{b}^\dag_{j,k}$ are the corresponding annihilation and creation operators of excitations at frequency $\omega_{j,k}$, which satisfy the canonical commutation relations 
\begin{eqnarray}\label{eq:can}
\left[\widehat{b}_{j,k}, \widehat{b}_{j',k'}\right] = 0, \quad  \left[\widehat{b}_{j,k}, \widehat{b}^\dag_{j',k'}\right] = \delta_{j j'}\delta_{k k'}.
\end{eqnarray}
Furthermore, $\widehat{H}_{I,j}(t)$ is the interaction Hamiltonian between the open system and the bath $j$; the time dependence is encoded in the function $\lambda_{j}(t)$ acting as an overall multiplicative factor, which may be used to model the coupling and decoupling of the system and the bath.
The parameters $g_{j,k}$ are the coupling strengths between the system and each bosonic mode, while the open system interacts with each bath with possibly different operators $\widehat{A}_{S,j}$.
As the initial condition of the dynamics, we restrict to the factorized state
\begin{equation}
  \label{eq:rhoSB0}
  \rho_{SB}(0) = \rho_S(0) \bigotimes_j \rho_{B,j}(0),
\end{equation}
where the bosonic baths are initially in Gibbs states, possibly at different inverse temperatures $\beta_j$,
\begin{equation}\label{eq:thst}
   \rho_{B,j}(0)= \frac{\exp\left[- \beta_j \widehat{H}_{B,j} \right]}{Z_j} \qquad Z_j = \Tr\left\{\exp\left[- \beta_j \widehat{H}_{B,j} \right]\right\}.
\end{equation}
The thermal state of each bath can be further written as a tensor product of thermal states for each frequency, since the bosonic modes are not interacting.
The model fixed by the Hamiltonian in Eq.~\eqref{eq:mainh} with the initial state in Eq.~\eqref{eq:rhoSB0} is a paradigmatic model of thermal environments in open quantum systems~\cite{breuer2002theory}, which predicts both dissipation and decoherence effects.
It is a particular instance of an open system interacting with Gaussian environments, meaning that the initial state of the environment is Gaussian, its free Hamiltonian is quadratic, and the environmental interaction operators are linear in the creation and annihilation operators.

Now, further exploiting that the initial environmental state is stationary with respect to the free dynamics, has zero mean values with respect to the environmental interaction operators, and is a product state among the different baths, we have that the open system's dynamics is completely determined by the autocorrelation functions of the baths:
\begin{equation}\label{eq:cj}
 C_j(t) = \langle X_j(t) X_j(0) \rangle = \Tr_B\left\{\widehat{U}_{B,j}(t)^\dag \widehat{X}_j \widehat{U}_{B,j}(t) \widehat{X}_j \rho_{B,j}(0) \right\},
\end{equation}
where 
\begin{equation}\label{eq:xj}
\widehat{X}_j=\sum_k g_{j, k}\left(\widehat{b}_{j,k}+\widehat{b}^\dag_{j,k}\right)
\end{equation}
are the Hermitian bath operators that appear in the interaction Hamiltonian, while
$\widehat{U}_{B,j}(t)=\exp \left[ -i \widehat{H}_{B,j} t \right]$ fixes the \emph{free} environmental dynamics.
Introducing the spectral densities of the baths
\begin{equation}
    J_j(\omega) =  \sum_{k} g^2_{j,k} \delta(\omega-\omega'_{j,k}),
\end{equation}
the autocorrelation functions can be equivalently written as
\begin{equation}
  \label{eq:autocorr_thermal_bath}
  C_j(t) =  \int_0^\infty d \omega J_j(\omega) \left[ \coth\left(  \frac{\beta_j \omega}{2} \right) \cos (\omega t) - i \sin (\omega t)  \right] = \int_{-\infty}^\infty d \omega J_{\beta_j,j}(\omega) e^{-i \omega t} ,
\end{equation}
where the dependence on the bath's temperature is fully enclosed in the hyperbolic cotangent factor.
We have also introduced the ``thermalized'' spectral density $J_{\beta_j,j}(\omega) = \mathrm{sign}(\omega) J_j(|\omega|) [ 1 + \coth (\beta_j \omega /2) ] / 2$, i.e. the Fourier transform of the autocorrelation function, in which the frequency spans both positive and negative values~\cite{Tamascelli2019}.

\subsection{Average thermodynamic quantities}\label{sec:atq}

Let us now see how to specifically define average thermodynamic quantities by referring to the global system-environment Hamiltonian; in doing so, we will essentially follow the definitions of Refs.~\cite{Esposito2010,Kato2016,Wiedmann2020a,Liu2021a,Landi2021a,Wiedmann2020a}.

Given a global time-dependent Hamiltonian (in the Schr{\"o}dinger picture), its expectation value can change:
\begin{equation}
  \label{eq:energyconsTdep}
  \frac{d}{dt} \Tr\left\{ \rho_{SB}(t) \widehat{H}_{\mathrm{tot}}(t)  \right\}  = \Tr\left\{ \rho_{SB}(t) \frac{\partial \widehat{H}_{\mathrm{tot}}(t) }{\partial t}  \right\} \equiv P_{\mathrm{tot}}(t),
\end{equation}
and a global energy change in a unitary transformation is considered as work, so that the right-hand side corresponds to the total power, which can be integrated on a generic time interval $[t_i, t_f]$ to obtain work during that interval.
The total work can be further split into work related to the system and to the interaction Hamiltonians:
\begin{equation}
  W_\mathrm{tot}(t_f,t_i) \equiv \int_{t_i}^{t_f} dt P(t) = \int_{t_i}^{t_f} dt \left( \Tr \left\{ \rho_{SB}(t) \frac{\partial \widehat{H}_{S}(t) }{\partial t}  \right\} + \Tr \left\{ \rho_{SB}(t) \frac{\partial \widehat{H}_{I}(t) }{\partial t}  \right\} \right) \equiv W_S(t_f,t_i) +  W_I(t_f,t_i),\label{eq:www}
\end{equation}
and the interaction work can be further split into contributions pertaining to each environment
\begin{equation}
  W_I(t_f,t_i) =  \sum_j \int_{t_i}^{t_f} dt \frac{d \lambda_j(t)}{dt} \Tr\left\{ \rho_{SB}(t) \left( \widehat{A}_{S,j}\otimes\sum_k g_{j, k}\left(\widehat{b}_{j,k}+\widehat{b}^\dag_{j,k}\right) \right) \right\} \equiv \sum_j^{N_B} W_{I,j}(t_f,t_i). \label{eq:wwi}
\end{equation}
Integrating the left-hand side of Eq.~\eqref{eq:energyconsTdep} we can split it into variations of internal energy of the open system, variations of the interaction energies and variations of the baths energies, which we interpret as heat, according to 
\begin{align}
  \label{eq:energy_variation}
  \int_{t_i}^{t_f} dt
  \frac{d}{dt} \Tr\left\{ \rho_{SB}(t) \widehat{H}_{\mathrm{tot}}(t)  \right\}&= U(t_f,t_i) + \sum_j^{N_B} \left[I_j(t_f,t_i) + Q_j(t_f,t_i)  \right] 
\end{align}
with
\begin{align}
  U(t_f,t_i) &= \Tr \left\{ \rho_{SB}(t_f)\widehat{H}_S(t_f) \right\}  - \Tr \left\{ \rho_{SB}(t_i)\widehat{H}_S(t_i) \right\} \label{eq:uu}\\
  I_j(t_f,t_i) & = \Tr \left\{ \rho_{SB}(t_f)\widehat{H}_{I,j}(t_f) \right\}  - \Tr \left\{ \rho_{SB}(t_i)\widehat{H}_{I,j}(t_i) \right\} \label{eq:ui}\\
  Q_j(t_f,t_i) &= \Tr \left\{ \left( \rho_{SB}(t_f) - \rho_{SB}(t_i) \right) \widehat{H}_B \right\}.\label{eq:uiq}
\end{align}
Energy conservation is thus written as 
\begin{equation}
  \label{eq:energyconservation}
  U(t_f,t_i) + \sum_j^{N_B} \left[ I_j(t_f,t_i) + Q_j(t_f,t_i) \right] = W_S(t_f,t_i) + \sum_j^{N_B} \left[ W_{I,j}(t_f,t_i) \right].
\end{equation}
We notice that in this setting we are taking the ``point of view'' of the total system (open system and baths), not of the external agent or system responsible for the time-dependence in the Hamiltonian.
Thus, the situation in which energy is being extracted from the system and baths in the form of work corresponds to a negative value of the instantaneous power $P_\mathrm{tot}(t)$, which means that the average energy of the total system is decreasing.

Furthermore, the exchanged heat is directly related to the entropy production, which quantifies the degree of irreversibility of the transformation experienced by the system.
In fact, in the case of an open quantum system interacting with $N_B$ bosonic baths initially in thermal states with temperatures $\beta_j$, $j=1,\ldots N_B$, see Eq.~\eqref{eq:thst}, the entropy production $\Sigma(t_f,t_i)$ in the time interval $[t_i, t_f]$ is given by~\cite{Esposito2010,Landi2021a}
\begin{equation}
\label{eq:Sigma_Esposito}
\Sigma(t_f,t_i) = \Delta S_S + \sum_{j=1}^{N_B} \beta_j Q_j(t_f,t_i),
\end{equation}
where
$\Delta S_S = S_S(t_f)-S_S(t_i)$ is the change in the system von Neumann entropy
\begin{equation}
S_S(t) =  - \Tr\left\{\rho_S(t) \ln \rho_S(t)\right\}.
\end{equation}

\subsection{Connection to correlation functions of the open-system}\label{sec:ctc}
The expressions derived so far allow us to deal with general dynamical regimes, including in principle arbitrary strong coupling and memory effects,
but involve global system-environment quantities.
Exploiting the linearity of the environmental interaction operators, it is however possible to define the same quantities in terms of two-time expectation values involving open-system observables only.
While the actual evaluation of multi-time (two-time) expectation values can be in principle as challenging as the evaluation of one-time expectation values of system-environment observables, this consideration is important from an operational point of view, since it relates the evolution of energy, heat and work
to quantities that can be accessed via measurements on the open-system only, along with the knowledge of the environmental correlation function.
Moreover, as discussed in Sec.~\ref{sec:tte}, these expressions can be useful when combined with an effective description of the environment
on a reduced set of degrees of freedom.
To simplify the notation, from this section onwards, we assume that $t_i=0$ and we denote all the quantities as dependent only on the final time $t_f$.

In Appendix~\ref{app:derivation_2time}, we exploit the Heisenberg picture of the global system-environment dynamics and a proper Bogoliubov transformation on the environmental modes derived from the framework of the thermofield dynamics~\cite{Landsman1987,Diosi1998,Yu2004,deVega2015b,Tamascelli2019},
to get the following expressions:
\begin{eqnarray}
  Q_j(t_f) &=& - 2 \int_0^{t_f} d t \lambda_j(t)
 \int_0^t ds \lambda_j(s) \Im\left\{ \dot{C}_j(t-s)
\left\langle A_{S,j}(t) A_{S,j}(s)\right\rangle\right\},\label{eq:qexpc}\\
W_{I,j}(t_f) &=& 2 \int_0^{t_f} d t \frac{d\lambda_j(t)}{d t}
 \int_0^t ds \lambda_j(s) \Im\left\{ C_j(t-s)
 \left\langle A_{S,j}(t) A_{S,j}(s)\right\rangle\right\}, \label{eq:wexpc}\\ 
 I_j(t_f ) &=& 2 \lambda_j(t_f) \int_0^{t_f} ds \lambda_j(s) \Im\left\{ C_j(t-s)  \left\langle A_{S,j}(t) A_{S,j}(s)\right\rangle \right\},\label{eq:iexpc}
\end{eqnarray}
where the dot denotes the time-derivative function $\dot{C} (t) = d C(t)/dt$, and $\left\langle A_{S,j}(t) A_{S,j}(s)\right\rangle$ denotes the two-time expectation value of the open-system interaction operator $\widehat{A}_{S,j}$, defined as \cite{Gardiner1985,breuer2002theory,Vacchini2024}
\begin{equation}
\left\langle A_{S,j}(t) A_{S,j}(s)\right\rangle = \Tr\left\{\widehat{U}_{tot}^\dag(t)\widehat{A}_{S,j}\otimes \mathbbm{1}_B\widehat{U}_{tot}(t)\widehat{U}_{tot}^\dag(s)\widehat{A}_{S,j}\otimes \mathbbm{1}_B\widehat{U}_{tot}(s)\rho_{SB}(0)\right\},
\label{eq:twotimemain}
\end{equation}
with the global unitary evolution operator
\begin{equation}
     \widehat{U}_{tot}(t) = T_{\leftarrow} e^{- i \int_{0}^t d \tau \widehat{H}_{tot}(\tau)},
\end{equation}
with $T_{\leftarrow}$ the time-ordering operator.
Analogous expressions have been derived previously, employing a generating functional approach~\cite{Kato2016,Gribben2022}.

\section{Thermodynamic quantities from system-pseudomode dynamics}
\label{sec:pseudomode}

As anticipated in the introduction, we are going to exploit a nonperturbative method for the open-system evolution, which is based on the replacement of the physical bath with a simpler effective environment composed of dissipative surrogate oscillators, the pseudomodes~\cite{Garraway1997,Tamascelli2017,Mascherpa2019}.

Given an open quantum system interacting with a Gaussian environment, being it bosonic or fermionic, the pseudomode approach allows one to replace
the original environment, typically made up of a continuous infinity of modes, via an auxiliary network of (possibly few) modes, which are in turn damped
by a Lindblad equation.
If the autocorrelation functions of such a network match the autocorrelation functions in Eq.~\eqref{eq:cj}, the reduced dynamics of the two configurations will be the same by virtue of the theorem proved in Ref.~\cite{Tamascelli2017} for bosonic environments and in Ref.~\cite{Chen2019g} for fermionic ones; note that a generalization of the approach via the use of non-Hermitian Hamiltonians has been introduced~\cite{Lambert2019a} and recently applied to the study of quantum thermodynamics~\cite{Menczel2024}.
In general, the autocorrelation functions in the two configurations will not be exactly the same and this will introduce a discrepancy in the corresponding reduced dynamics, which however can be kept under control by means of general bounds~\cite{Mascherpa2016} or by checking the numerical convergence as the number of auxiliary modes increases.

\subsection{The pseudomode setup}\label{sec:tpa}

The evolution of the open system interacting with the network of damped pseudomodes is fixed by the Lindblad equation
\begin{equation}\label{eq:ddt}
    \frac{d}{dt}\rho_{SM}(t) = \mathcal{L}_{SM}(t)[\rho_{SM}(t)],
\end{equation}
where $\rho_{SM}(t)$ denotes the system-pseudomodes state at time $t$, while $\mathcal{L}_{SM}(t)$ is the generator
\begin{equation}\label{eq:lpseu}
\mathcal{L}_{SM}(t)[\rho_{SM}] = - i\left[\widehat{H}_S(t)+ \sum_{j=1}^{N_B}\left(\widehat{H}_{M,j}+
\widehat{H}_{IM,j}(t)\right),
\rho_{SM}\right] + \sum_{j=1}^{N_B} \sum^{n_j}_{k=1} \mathcal{D}_{j,k}[\rho_{SM}],
\end{equation}
where we have introduced a different set of $n_j$ pseudomodes for each of the $N_B$ baths.
The system Hamiltonian $\widehat{H}_S(t)$ is the same as the one in Eq.~\eqref{eq:mainh}, the interaction Hamiltonian between the open system and the pseudomodes is
\begin{equation}\label{eq:himj}
\widehat{H}_{IM,j}(t) = \lambda_j(t)\widehat{A}_{S,j}
\otimes \sum_{k=1}^{n_j} \sqrt{c_{j,k}} \left(\widehat{a}^\dag_{j,k}+\widehat{a}_{j,k}\right),
\end{equation}
where the time-dependent function $\lambda_j(t)$ and the open-system interaction operator $\widehat{A}_{S,j}$ are the same as those of the original interaction Hamiltonian in Eq.~\eqref{eq:mainh}, and $\widehat{a}_{j,k}$ and $\widehat{a}^\dag_{j,k}$ are the annihilation and creation operators of the $k$-th pseudomode associated with the $j$-th bath.

In the following, we will focus on two kinds of networks of damped harmonic oscillators: nearest-neighbor interacting pseudomodes at zero temperature or non-interacting finite-temperature ones.
In the former case, the free Hamiltonian of the pseudomodes of the $j$-th bath is
\begin{equation}
\widehat{H}_{M,j} = 
\sum_{k=1}^{n_j} \Omega_{j,k} \widehat{a}^\dag_{j,k}\widehat{a}_{j,k} 
+ \sum_{k=1}^{n_j-1} \left(g_{j,k} \widehat{a}^\dag_{j,k}\widehat{a}_{j,k+1}+ g^*_{j,k} \widehat{a}_{j,k}\widehat{a}^\dag_{j,k+1}\right),
\end{equation}
where we have separated the free Hamiltonians of each pseudomode and the interacting terms among the pseudomodes, while the dissipative term acting on the $k$-th pseudomode of the $j$-th bath is
\begin{eqnarray}
\mathcal{D}_{j,k}[\rho_{SM}] &=& \Gamma_{j,k}\left(\widehat{a}_{j,k} \rho_{SM} \widehat{a}^\dag_{j,k}
-\frac{1}{2}\left\{\widehat{a}^\dag_{j,k}\widehat{a}_{j,k}, \rho_{SM}\right\}\right),\label{eq:disspz}
\end{eqnarray}
where $\Gamma_{j,k} \geq 0$ are the damping rates.
Finally, the initial system-pseudomode state is a product state 
$\rho_{SM}(0) = \rho_S(0)\otimes \rho_M(0)$, with 
\begin{equation}
\rho_M(0) = \bigotimes_{j=1}^{N_B} \bigotimes_{k=1}^{n_j}\ket{0}_{j,k}\bra{0}_{j,k}
\end{equation}
the vacuum of the pseudomodes.

For the case of non-interacting pseudomodes at finite temperature, their free Hamiltonian is indeed of the form
\begin{equation}\label{eq:hmj}
\widehat{H}_{M,j} = 
\sum_{k=1}^{n_j} \Omega_{j,k} \widehat{a}^\dag_{j,k}\widehat{a}_{j,k},
\end{equation}
while the dissipator now reads
\begin{eqnarray}
\mathcal{D}_{j,k}[\rho_{SM}] &=& \Gamma_{j,k}(1+\bar{n}_{j,k})\left(\widehat{a}_{j,k} \rho_{SM} \widehat{a}^\dag_{j,k}
-\frac{1}{2}\left\{\widehat{a}^\dag_{j,k}\widehat{a}_{j,k}, \rho_{SM}\right\}\right)\nonumber\\
&&+\Gamma_{j,k} \bar{n}_{j,k}\left(\widehat{a}^\dag_{j,k} \rho_{SM} \widehat{a}_{j,k}
-\frac{1}{2}\left\{\widehat{a}_{j,k}\widehat{a}^\dag_{j,k}, \rho_{SM}\right\}\right),\label{eq:dissp}
\end{eqnarray}
where (using units such that $k_B = 1$ so that $\beta_{j,k}=T_{j,k}^{-1}$)
\begin{equation}\label{eq:njk}
\bar{n}_{j,k} = \frac{1}{e^{\Omega_{j,k}/T_{j,k}} -1}
\end{equation}
is the thermal occupation number of the pseudomode at frequency $\Omega_{j,k}$, given its temperature $T_{j,k}$.
The initial system-pseudomode state is now the product state $ \rho_{SM}(0) = \rho_S(0)\otimes \rho_M(0)$, with the pseudomodes initially in the thermal state 
\begin{equation}
\rho_M(0) = \bigotimes_{j=1}^{N_B} \bigotimes_{k=1}^{n_j} \rho^\beta_{j,k}, \qquad \rho^\beta_{j,k} = \frac{e^{-\widehat{H}_{M,j}/T_{j,k}}}{\Tr\left\{e^{-\widehat{H}_{M,j}/T_{j,k}}\right\}}.
\end{equation}

In both cases, the expectation values of the pseudomode interaction operators on the initial state are zero, and the initial state is stationary with respect to the free dynamics, so that the reduced dynamics of the open-system is fixed by the pseudomode autocorrelation functions
\begin{equation}\label{eq:cmjj}
    C_{M,j}(t) = \Tr_M\left\{\widehat{X}_{M,j} e^{\mathcal{L}_{M,j} t} \left[\widehat{X}_{M,j} \rho_{M,j}(0)\right] \right\},
\end{equation}
where we introduced the pseudomode interaction operators appearing in Eq.~\eqref{eq:himj}
\begin{equation}\label{eq:xmj}
    \widehat{X}_{M,j} = \sum_{k=1}^{n_j} \sqrt{c_{j,k}} \left(\widehat{a}^\dag_{j,k}+\widehat{a}_{j,k}\right)
\end{equation}
and the Lindblad generators
\begin{equation}
\mathcal{L}_{M,j}\left[\rho_{M,j}\right] = 
- i\left[\widehat{H}_{M,j},
\rho_{M,j}\right] + \sum^{n_j}_{k=1} \mathcal{D}_{j,k}[\rho_{M,j}]\label{eq:lgs}
\end{equation}
that act on the pseudomodes associated with the $j$-th bath only and determine their free dynamics, i.e., their dynamics when decoupled from the open system (and then also from the other baths, since there is no direct interaction among them and the initial state is factorized also with respect to the $N_B$ baths).
The equality
\begin{equation}\label{eq:eqq}
   C_j(t) = C_{M,j}(t) \quad \forall t\geq 0, \,\, \forall j
\end{equation}
guarantees the equality of the open-system dynamics, respectively, in the original unitary configuration fixed by the Hamiltonian Eq.~\eqref{eq:mainh},
and in the configuration with the network of pseudomodes fixed by Eqs.~\eqref{eq:ddt} and~\eqref{eq:lpseu}.

The key point is then of course to find appropriate parameters for the network of damped modes so that $C_{M,j}(t)$ approximates $C_j(t)$ as well as possible and Eq.~\eqref{eq:eqq} approximately holds, for any given number of pseudomodes.
In the case of $n_j$ possibly interacting zero-temperature pseudomodes, the autocorrelation functions is of the form~\cite{Mascherpa2019}
\begin{equation}\label{eq:cmj2}
    C_{M,j}(t) = \sum_{k=1}^{n_j} w_{j,k} e^{\chi_{j,k} t},
\end{equation}
as follows from the semigroup evolution in Eq.~\eqref{eq:cmjj}.
The coefficients $w_{j,k}$ and $\chi_{j,k}$ depend nonlinearly on the parameters fixing the network of damped harmonic oscillators, i.e., the free frequencies $\Omega_{j,k}$, the couplings among the pseudomodes $g_{j,k}$, the couplings with the open system $c_{j,k}$, and the damping rates $\Gamma_{j,k}$.
Starting from a generic unitary autocorrelation function $C_{j}(t)$, one can then first perform an exponential fit of the latter, $\tilde{C}_j(t) = \sum_{k=1}^{n_j} \tilde{w}_{j,k} e^{\tilde{\chi}_{j,k} t}$---such that $\tilde{C}_j(t)$ approximates $C_j(t)$ as well as possible---and then determine the coefficients of the pseudomodes so that $w_{j,k}$ and $\chi_{j,k}$ approximate $\tilde{w}_{j,k}$ and $e^{\tilde{\chi}_{j,k} t}$ as well as possible.
Remarkably, in the case of interacting pseudomodes the coefficients $w_{j,k}$ can be complex numbers, which can significantly reduce the number of exponentials needed to fit a given correlation function.
A procedure that can deal with a general configuration of the pseudomodes by solving a nonlinear inversion problem has been developed in Ref.~\cite{Mascherpa2019}; in particular, when dealing with zero-temperature interacting pseudomodes, we will consider unitary autocorrelation functions that have been already treated there and whose set of auxiliary pseudomode coefficients are thus already known.

The procedure to determine the auxiliary network of pseudomodes simplifies considerably if one restricts to non-interacting pseudomodes.
For $n_j$ pseudomodes at temperatures $T_{j,k}$, one has~\cite{Somoza2019}
\begin{equation}\label{eq:cmj3}
    C_{M,j}(t) = \sum_{k=1}^{n_j} c_{j,k} e^{-\frac{\Gamma_{j,k}}{2} t}
    \left(\coth\left(\frac{\Omega_{j,k}}{2 T_{j,k}}\right)\cos\left(\Omega_{j,k} t\right) - i \sin\left(\Omega_{j,k} t\right)  \right),
\end{equation}
which is directly expressed in terms of the parameters of the pseudomodes configuration.
The latter can be thus easily determined by matching $C_j(t)$ with the expression in Eq.~\eqref{eq:cmj3}.

\subsubsection{Multi-time quantities}
Recently~\cite{Smirne2022a}, it has been shown that the pseudomode approach can be employed also to describe the multitime correlation functions 
of open-system quantities.
We briefly recall here the general result, which will be later applied to get a unified framework to treat the multi-time open-system expressions of the thermodynamic quantities presented in Sec.~\ref{sec:ctc}.

The most general multi-time expectation value for an open quantum system reads~\cite{Accardi1982,Dumcke1983,Li2018f}
\begin{eqnarray}
&&\langle E_n(t_n) \ldots E_1(t_1); F_1(t_1)\ldots F_n(t_n)\rangle = 
\Tr_{SE}\left\{\widehat{E}_n(t_n)
\ldots \widehat{E}_1(t_1)\rho_{SB}(0)
\widehat{F}_1(t_1)\ldots\widehat{F}_n(t_n)
\right\},\label{eq:multi}
\end{eqnarray}
where $\rho_{SB}(0)$ is the initial system-environment state, the time dependence of the operators is indeed in the Heisenberg picture,
\begin{equation}\label{eq:eut}
    \widehat{E}_k(t_k) = \widehat{U}^\dag_{tot}(t_k) \widehat{E}_k\otimes \mathbbm{1}_B \widehat{U}_{tot}(t_k),
\end{equation}
and analogously for $\widehat{F}_k(t_k)$, where $\widehat{E}_k\otimes \mathbbm{1}_B$ and $\widehat{F}_k\otimes\mathbbm{1}_B$ are generic open-system operators; we will always refer to time-ordered expectation values, i.e., with $t_n\geq \ldots \geq t_1\geq 0$.
If the operators on the right of the initial state are the adjoint of those on the left, $\widehat{F}_k = \widehat{E}^\dag_k$ for any $k$, we recover the multi-time probabilities associated with a sequence of generic measurements~\cite{Milz2020}, while if the right-side operators are equal to the identity, $\widehat{F}_k=\mathbbm{1}_S$, we recover the multi-time correlation functions of the left-side observables, which are associated with perturbative expansions~\cite{breuer2002theory}, as well as with the expressions of the thermodynamics quantities in Eq.~\eqref{eq:twotimemain}.

Under the same assumptions that lead to the equivalence of the reduced dynamics, and in particular under the validity of Eq.~\eqref{eq:eqq}, the open-system multi-time expectation values in Eq.~\eqref{eq:multi} are equal to those in the auxiliary pseudomode configuration fixed by the generator $\mathcal{L}_{SM}$ in Eq.~\eqref{eq:lpseu}, which are given by
\begin{eqnarray}
&&\langle E_n(t_n) \ldots E_1(t_1); F_1(t_1)\ldots F_n(t_n)\rangle_L =  \label{eq:ooL}\\
&&
\hspace{2.cm}\Tr_{SM}\left\{\widehat{E}_n\otimes \mathbbm{1}_M \Lambda_{SM}(t_n,t_{n-1})\left[\ldots 
\Lambda_{SB}(t_2,t_1)\left[\widehat{E}_1\otimes \mathbbm{1}_M\Lambda_{SM}(t_1)\left[\rho_{SM}(0)\right]
\widehat{F}_1\otimes \mathbbm{1}_M\right]\ldots
\right]
\widehat{F}_n\otimes \mathbbm{1}_M \right\},\notag
\end{eqnarray}
where $\rho_{SM}(0)$ is the initial system-pseudomode state, and we introduced the system-pseudomode propagators
\begin{equation}\label{eq:lsm}
\Lambda_{SM}(t,s) = T_{\leftarrow}e^{\int_{s}^t\mathcal{L}_{SM}(s)},
\end{equation}
with $\Lambda_{SM}(t)=\Lambda_{SM}(t,0)$. In other terms, by introducing the pseudomodes, one extends the set of considered degrees of freedom in a way such that, not only the dynamics can be evaluated via a Lindblad equation on the extended configuration, but also the multi-time expectation values are given by the quantum regression formula~\cite{Lax1968,Gardiner1985,breuer2002theory}, i.e., they are fixed by the same propagators that determine the dynamics 
(compare Eqs.~\eqref{eq:multi} and~\eqref{eq:eut} with Eqs.~\eqref{eq:ooL} and~\eqref{eq:lsm}).

\subsubsection{Product of environmental interaction operators}
The transition from the original unitary configuration to the pseudomodes reduces considerably the number of degrees of freedom that are relevant to determine the time evolution of open-system observables.
However, since it is not an overall unitary transformation, there is a priori no inherent connection between quantities associated with the original bath and those associated with the pseudomode environment.

As recently proved in Ref.~\cite{Menczel2024}, an important exception to this is represented by the expectation values of  tensor products of system operators and strings of environmental interaction operators.
Explicitly, under the same assumptions of the equivalence theorem for the reduced dynamics, in particular Eq.~\eqref{eq:eqq}, one has the identity
\begin{equation}\label{eq:eqpre}
\Tr_{SB}\left[\widehat{O}_S\otimes \widehat{X}_{j_1}\ldots\widehat{X}_{j_k} \rho_{SB}(t) \right]
=\Tr_{SM}\left[\widehat{O}_S\otimes \widehat{X}_{M,j_1}\ldots\widehat{X}_{M,j_k} \rho_{SM}(t) \right],
\end{equation}
where $\widehat{O}_S$ is a generic open-system operator, $\widehat{X}_{j_1}, \ldots,\widehat{X}_{j_k}$ is a sequence of $k$ environmental interaction operators of the unitary configuration, as defined in Eq.~\eqref{eq:xj}, and $\widehat{X}_{M,j_1}\ldots,\widehat{X}_{M,j_k}$ is the same sequence of the $k$ corresponding interaction operators for the pseudomode configuration, see Eq.~\eqref{eq:xmj}.
Note that $\widehat{X}_{j}$ and $\widehat{X}_{M,j}$ are in general different operators, but they refer to the same $j$-th bath, so that the sequence $j_1, \ldots, j_k$ univocally determines the involved environmental interaction operators in both configurations.

In Appendix~\ref{app:pef} we provide an alternative proof of the statement above, which goes along the same lines as the proof of the equivalence theorem in Ref.~\cite{Tamascelli2017} and is essentially based on considering the global evolution in Heisenberg picture. 
In Sec.~\ref{subsec:onetimeexp}, we will show how such a relation establishes a direct connection between the thermodynamics quantities we are investigating and the expectation values of appropriate global system-pseudomode quantities.

\subsection{Two-time expressions}\label{sec:tte}
The most direct way to exploit pseudomodes to evaluate the thermodynamic quantities of heat, interaction work and interaction energy is indeed to refer to their expressions in Eqs.~\eqref{eq:qexpc}--\eqref{eq:iexpc}, which involve the environmental autocorrelation functions $C_j(t)$---that are supposed to be known---along with the two-time correlation functions of the open-system interaction operators $\left\langle A_{S,j}(t)A_{S,j}(s)\right\rangle$.
The latter are in fact special instances of the general expression in Eq.~\eqref{eq:multi}, for $n=2$, $\widehat{E}_1 = \widehat{E}_2 = \widehat{A}_{S,j}$
and $\widehat{F}_1 = \widehat{F}_2 = \mathbbm{1}_S$, so that if Eq.~\eqref{eq:eqq} holds, we can evaluate the average thermodynamic quantities
in Eqs.~\eqref{eq:qexpc}--\eqref{eq:iexpc} by using the two-time expression of $\left\langle A_{S,j}(t)A_{S,j}(s)\right\rangle$ provided by the pseudomode configuration, i.e.,
\begin{equation}\label{eq:twotimeps}
\left\langle A_S(t)A_S(s)\right\rangle = 
\Tr_{SM}\left\{\widehat{A}_S \Lambda_{SM}(t,s)\left[
\widehat{A}_S \Lambda_{SM}(s,0) [\rho_{SM}(0)]\right] \right\}.
\end{equation}
In principle, knowledge of the propagator $\Lambda_{SM}(t,s)$ would allow to evaluate all the relevant quantities.
Concretely, the expression in Eq.~\eqref{eq:lsm} corresponds to a numerical integration of the time-dependent Lindblad equation $\frac{d}{dt} \Lambda_{SM}(t,s) = \mathcal{L}_{SM}(t) \Lambda_{SM}(t,s)$, starting from $t=s$ with the initial condition $\Lambda_{SM}(s,s)=\mathcal{I}$ (the identity superoperator).
This approach is quite computationally intensive: if $d$ is the dimension of the overall Hilbert space of system and pseudomodes, then $\Lambda_{SM}(t,s)$ corresponds to a $d^2{\times}d^2$ matrix, which quickly becomes hard to handle numerically.

A more efficient and standard approach to evaluate the two-time expression in Eq.~\eqref{eq:twotimeps} is to first integrate the Lindblad equation for $\rho_{SM}(s)$, i.e. using $\rho_{SM}(0)$ as the initial condition, and then integrate the same equation from $s$ to $t$ using $\widehat{A}_S \rho_{SM}(s)$ as the initial condition.
Thus, one can compute the expectation values by solving multiple times the Lindblad equations for states, without dealing with the propagator 
$\Lambda_{SM}(t,s)$, which allows working with larger Hilbert space dimensions.
A third approach is to use the Monte Carlo wave-function method, see e.g.~\cite[Sec.~3.2.2]{Daley2014}. 
This is certainly convenient in terms of memory, since the computation handles with $d$-dimensional vectors instead of $d{\times}d$ density matrices; however this procedure requires a large number of samples to ensure convergence of all the quantities.
Regardless of how the values of $\left\langle A_S(t)A_S(s)\right\rangle$ are obtained, they must then be inserted in the integrals over $(t,s)$ that define the thermodynamic quantities of interest in Eqs.~\eqref{eq:qexpc}--\eqref{eq:iexpc}.
The most straight-forward approach is to approximate such integrals with finite sums (e.g. with the trapezoidal rule).
For this reason, one needs to obtain the values on a $(t,s)$ grid fine enough so that the numerical error of this last integration step can be negligible; this may pose an additional challenge to get numerically accurate results.

\subsection{One-time expressions}\label{subsec:onetimeexp}
Exploiting the correspondence between the expectation values involving a generic sequence of environmental interaction operators, one can derive expressions of the average thermodynamic quantities that only involve one-time system pseudomode quantities, following what has been done in Ref.~\cite{Menczel2024} for the exchanged heat.

In fact, taking the unitary expression for the interaction work in Eq.~\eqref{eq:wwi}---for $t_i=0$---and using Eq~\eqref{eq:xj}, we have 
\begin{eqnarray}
W_{I,j}(t_f) &=& \int_{0}^{t_f} dt \frac{d \lambda_j(t)}{dt} \Tr_{SB}\left\{ \rho_{SB}(t) \left( \widehat{A}_{S,j}\otimes \widehat{X}_j\right) \right\},  \label{eq:prewexpc_pseudo} 
\end{eqnarray}
so that Eq.~\eqref{eq:eqpre} for $k=1$ and $\widehat{O}_{S}=\frac{d \lambda_j(t)}{dt}\widehat{A}_{S,j}$, along with Eqs.~\eqref{eq:himj} and \eqref{eq:xmj}, 
lead us to
\begin{eqnarray}
W_{I,j}(t_f) &=&  \int_0^{t_f} dt \Tr_{SM}\left\{  \frac{\partial \widehat{H}_{IM,j}(t)}{\partial t} \,  \rho_{SM}(t) \right\},  \label{eq:wexpc_pseudo} 
\end{eqnarray}
where $\rho_{SM}(t)$ denotes the system-pseudomode state
and $\widehat{H}_{IM,j}(t)$ is the interaction Hamiltonian between the system and the pseudomodes associated with the $j$-th bath in Eq.~\eqref{eq:himj}.
Analogously, for the interaction energy in Eq.~\eqref{eq:ui} we directly get
\begin{eqnarray}
 I_j(t_f) &=& \Tr_{SM}\left\{  \widehat{H}_{IM,j}(t_f) \rho_{SM}(t_f)  \right\}. \label{eq:ipc_pseudo}
\end{eqnarray}

Crucially, we can apply the equivalence in Eq.~\eqref{eq:eqpre} also to the expression for the heat, by rewriting it according to elementary manipulations (see Eq.~\eqref{eq:qjtau2}) in the form
\begin{eqnarray}
Q_j(t_f) = i \int_0^{t_f} d t \Tr\left\{ \widehat{H}_{I,j}(t) \left[ \widehat{H}_{B,j},\rho_{SB}(t)\right]\right\}\label{eq:extraq} 
\end{eqnarray}
so that we have
\begin{eqnarray}
&&Q_j(t_f) = 
i \int_0^{t_f} d t \Tr\left\{\left[\widehat{H}_{I,j}(t), \widehat{H}_{tot}(t)-\widehat{H}_{S}(t) -\sum_{j'=1}^{N_B}\widehat{H}_{I,j'}(t)\right] \rho_{SB}(t)\right\}\nonumber\\
&=& -i \int_0^{t_f} d t\left(\Tr_{SB}\left\{\left[\widehat{H}_{I,j}(t),\widehat{H}_{S}(t)+\sum_{j'=1}^{N_B}\widehat{H}_{I,j'}(t)\right] \rho_{SB}(t)\right\}
- \frac{d}{dt}\Tr\left\{\widehat{H}_{I,j}(t)\rho_{SB}(t)\right\}+\Tr\left\{\frac{\partial\widehat{H}_{I,j}(t)}{\partial t}\rho_{SB}(t)\right\}\right) \nonumber \\ 
&=& -i \int_0^{t_f} d t\left(\Tr_{SM}\left\{\left[\widehat{H}_{IM,j}(t),\widehat{H}_{S}(t)+\sum_{j'=1}^{N_B}\widehat{H}_{IM,j'}(t)\right] \rho_{SM}(t)\right\}
- \Tr\left\{\widehat{H}_{IM,j}(t) \frac{d}{dt} \rho_{SM}(t)\right\} \right) .\label{eq:extraq} 
\end{eqnarray}
In the second line we have explicitly expressed heat in terms of expectation values of system and interaction operators only; for the third line we have used Eq.~\eqref{eq:eqpre} to replace system-bath quantities with system-pseudomodes ones.
Note that since the equivalence holds for any $t$, it must hold also for the time-derivative of the expectation value. 
Using the equation of motion for the pseudomodes in Eqs.~\eqref{eq:ddt} and~\eqref{eq:lpseu}, one gets~\cite{Menczel2024}
\begin{eqnarray}
  Q_j(t_f) &=& - \int_0^{t_f} dt \Tr\left\{ \widehat{H}_{IM,j}(t) \, \mathcal{L}_{M,j} [ \rho_{SM}(t)  ]\right\},\label{eq:qexpc_pseudo} 
\end{eqnarray}
where $\mathcal{L}_{M,j}$ is the generator of the pseudomodes' free dissipative dynamics in Eq.~\eqref{eq:lgs}.
We notice that this final expression is formally equivalent to the one for the original bath, cf. the first equality in Eq.~\eqref{eq:extraq}, since $-i [ \widehat{H}_{B,j},\rho_{SB}(t) ]$ is the generator of the bath's free unitary dynamics applied to the system and bath state.
We conclude that the calculation in the pseudomode picture amounts to replacing the original environment and the (Hamiltonian) generator of its free dynamics with the pseudomodes and their (Linbdladian) generator.
We reiterate that this is possible because heat can be expressed in terms of open-system and environmental-interaction operators only.
We stress that the expressions in Eqs.~\eqref{eq:wexpc_pseudo}, \eqref{eq:ipc_pseudo} and \eqref{eq:qexpc_pseudo} only involve single-time expressions, and then one-time integrals, so that their explicit evaluation is considerably less expensive than the two-time expressions in Eqs.~\eqref{eq:qexpc}--\eqref{eq:iexpc}.
We will therefore rely on this approach in the following analysis of concrete examples.

\section{Examples}
\label{sec:examples}

In this section we apply the pseudomode framework to calculate thermodynamic quantities presented in the previous section to a couple of concrete exemplary problems.
In both cases, we consider a two-level system (TLS), i.e. the well-known spin-boson model.
More specifically, we choose a free Hamiltonian $\widehat{H}_S =  \omega_0 \frac{\sigma_z}{2} $ and a coupling operator $\widehat{A}_S = \sigma_x$.
As the first application, we study the entropy production of a TLS in contact with a single bath with an ohmic spectral density.
In the second example we consider the TLS in contact with two thermal baths, studying a simple example of a thermal machine obtained by modulating the coupling between the TLS and the cold bath.

\subsection{Entropy production of a two-level system dissipating in a structured environment}

We consider a TLS in contact with a single thermal bath and the main quantity of interest is the entropy production.
This quantity is defined in Eq.~\eqref{eq:Sigma_Esposito} and, in principle, requires knowledge of the expectation value of a bath observable, thus we at least some insight into the \emph{global} dynamics of system and bath.
Since it was introduced by Esposito, Lindenberg and van den Broeck (ELB)~\cite{Esposito2010}, this definition of entropy production will be denoted as $\Sigma_{\mathrm{ELB}}$ in this section.
For a single bath it reads: 
\begin{equation}
  \Sigma_{\mathrm{ELB}}(t) = \Delta S (t) + \beta Q(t) =  \Delta S (t) + \beta \left[  -U(t) - I(t) + W_S(t) + W_I(t)   \right],
  \label{eq:ELB_example}
\end{equation}
where the second equality is due to energy conservation as expressed in Eq.~\eqref{eq:energyconservation}.

For comparison, we also consider a definition solely based on the \emph{local} reduced dynamics of the open system, originally introduced by Spohn~\cite{Spohn1978} for time-independent Lindblad dynamics:
\begin{align}
  \Sigma_{\mathrm{Spohn}}(t) &= S( \rho_S(0)  || \rho_{S,\mathrm{Gibbs}} ) - S( \rho_S(t)  || \rho_{S,\mathrm{Gibbs}} ) = \Delta S (t) - \beta U(t),
  \label{eq:Spohn_example}
\end{align}
where $\rho_{S,\mathrm{Gibbs}} = e^{-\beta H_S} / \Tr \left[ e^{-\beta H_S} \right] $ is the Gibbs thermal state of the system.
We notice that there are no time-dependent terms in the Hamiltonian, thus we have $W_S(t)=0=W_I(t)$ and the two definitions match when $Q(t) = -U(t)$, i.e. when the interaction energy is equal to zero.
Even though this generally does not happen exactly, there are situations of interest where the interaction energy can be neglected, as is usually the case in classical thermodynamics for macroscopic systems.

This equation was generalized to the case of a time-dependent system Hamiltonian $\widehat{H}_S(t)$ by Deffner and Lutz~\cite{Deffner2011} 
\begin{align}
  \Sigma_{\mathrm{DL}}(t) &= 
  S( \rho_S(0)  || \rho_{S,\mathrm{Gibbs}}(0) ) - S( \rho_S(t)  || \rho_{S,\mathrm{Gibbs}}(t) ) - \int_0^t ds  \Tr \left[ \rho_S(s) \partial_s \log \rho_{S,\mathrm{Gibbs}}(s)   \right] \\ 
  &= \Delta S (t) - \beta U(t) + \beta W_S(t),
\end{align}
where $\rho_{S,\mathrm{Gibbs}}(t) = e^{-\beta H_S(t)} / \Tr \left[ e^{-\beta H_S(t)} \right]$ is the instantaneous Gibbs state of the time-dependent system Hamiltonian. 
We stress that, despite the appearance of the instantaneous Gibbs state in this expression, there are no assumptions of fast relaxation, and it is just a convenient way to rewrite the familiar thermodynamic expression in the second line.
Similarly to the time-independent case, this definition corresponds to the global one when the contribution due to the interaction is equal to zero, i.e. when $\beta( - I(t) + W_I(t) ) = 0$.

Related studies include Refs.~\cite{Pucci2013,Colla2021} for a harmonic oscillator system, by direct access to the bath of harmonic oscillators and Ref.~\cite{Goyal2020}, which studied two spins in contact with a bosonic bath with hierarchical equations of motion.

\subsubsection{Results}
\begin{figure}[t]
  \includegraphics{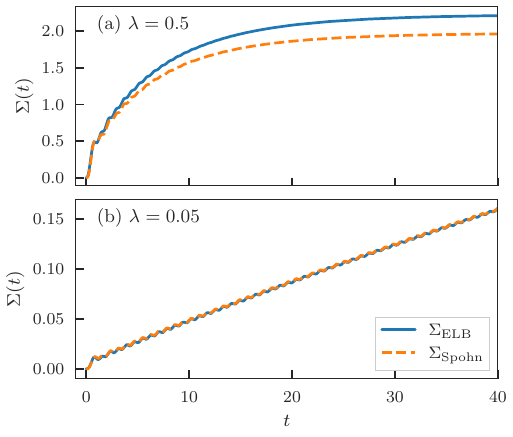}
  \caption{Comparison of the ELB and Spohn definitions of entropy production as a function of time, for an open TLS interacting with a bosonic thermal bath with ohmic spectral density with exponential cutoff at frequency $\Omega_c$.
  The temperature of the bath is set to $T=\Omega_c$, the frequency of the TLS is $\omega_0=2 \Omega_c$, time $t$ is in units of $\Omega_c^{-1}$.
  This figure shows the static case without any time-dependent driving.
  Panels (a) and (b) correspond to relatively strong and weak coupling, respectively.
  }
  \label{fig:static_entropyprod_dynamics}
\end{figure}

We show results for a thermal bath with an ohmic spectral density with exponential cutoff 
\begin{equation}
  J(\omega)=\pi \omega e^{-\omega / \Omega_c},
\end{equation}
where $\Omega_c$ is the cutoff frequency.
We fix the temperature to the particular value $T=\Omega_c$ and the frequency of the system Hamiltonian to $\omega_0 = 2 \Omega_c$.
In particular, we employ the parameters tabulated in Ref.~\cite{Mascherpa2019} to obtain the effective pseudomode description of the bath.
In Fig.~\ref{fig:static_entropyprod_dynamics} we show a comparison between the time evolution of the global entropy production~\eqref{eq:ELB_example} and the local definition~\eqref{eq:Spohn_example}. 
As the system-bath coupling strength $\lambda$ decreases, we can see that the two quantities get closer, as expected from the previous general arguments.

\begin{figure}[t]
  \includegraphics{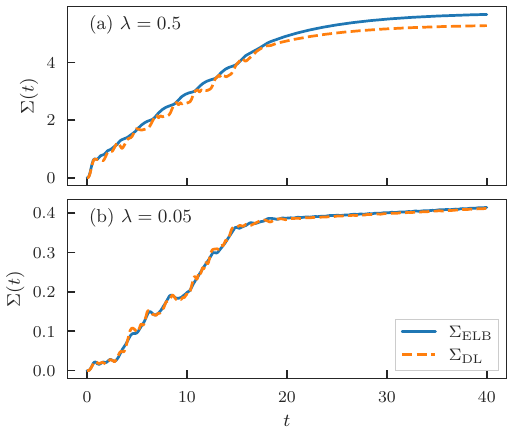}
  \caption{Comparison of the ELB and Deffner-Lutz definitions of entropy production as a function of time, for an open TLS interacting with a bosonic thermal bath with ohmic spectral density with exponential cutoff at frequency $\Omega_c$.
  The temperature of the bath is set to $T=\Omega_c$, the frequency of the TLS is $\omega_0=2 \Omega_c$, time $t$ is in units of $\Omega_c^{-1}$.
  The TLS is driven by the time-dependent Hamiltonian~\eqref{eq:HtdepTLS} with $F_0 = 10$, $\omega_f= \Omega_c$ and $\Omega_f = \Omega_c / 20$.
  Panels (a) and (b) correspond to relatively strong and weak coupling, respectively.
  }
  \label{fig:timedep_entropyprod_dynamics}
\end{figure}
We then consider the situation in which the system Hamiltonian contains a time-dependent part
\begin{equation}
  \label{eq:HtdepTLS}
  \widehat{H}_S(t) = \widehat{H}_0 + f(t) \sigma_x,     
\end{equation}
with the following choice of a sinusoidal driving~\cite{Colla2021} of duration $T_f=\pi/\Omega_f$:
\begin{equation}
  f(t) = \begin{cases} 
    F_0 \sin ( \omega_f t ) \sin^2 ( \Omega_f t ) \qquad & t \leq \frac{\pi}{\Omega_f} \\ 
    0 & t > \frac{\pi}{\Omega_f}.
  \end{cases}
\end{equation}
Keeping the same parameters of Fig.~\ref{fig:static_entropyprod_dynamics}, and further choosing $F_0 = 10$ and $T_f = 20 / \Omega_c$ and $\omega_f = \Omega_c$, in Fig.~\ref{fig:timedep_entropyprod_dynamics} we show again a comparison between the dynamics of the global and local entropy production definitions.
Also in this case, we see that as the system-bath coupling strength $\lambda$ decreases the two quantities get closer.
An interesting open research question is thus to understand the conditions under which the two definitions of entropy production coincide in the weak coupling limit, from general analytical arguments and beyond numerical studies of particular models.

\subsection{Thermal machine with sinusoidally-modulated system-bath coupling}

A minimal setup to obtain a heat engine that extracts useful work from two baths consists of a harmonic modulation of one the two system-bath couplings, under some conditions on the spectral densities and frequency of the modulation~\cite{Cavaliere2022}.
This behaviour was shown for a harmonic oscillator coupled to two bosonic baths, for which analytical or semi-analytical solutions are available~\cite{Freitas2016,Cavaliere2022}.
We investigate the same protocol numerically using our nonperturbative framework when the role of the thermodynamic ``working substance'' connecting the two baths is played by a TLS instead of a harmonic oscillator.
We remark that the interaction Hamiltonian between the bath and the harmonic oscillator system assumed in Ref.~\cite{Cavaliere2022} contains an explicit time-dependent Hamiltonian term acting only on the system, needed to ensure the translational invariance~\cite{Weiss2012}.
Since here we consider a TLS, this term is absent~\cite{Leggett1987}.

We work in the following configuration: the system interacts with a cold structured bath that has a spectral density peaked at a particular frequency $\omega_1$ (a Lorentzian peak was chosen in Ref.~\cite{Cavaliere2022}) and this interaction is modulated by a sinusoidal function (i.e. a harmonic drive).
Simultaneously, the system interacts with a hot bath without prominent spectral features (chosen to be ohmic in Ref.~\cite{Cavaliere2022}); this interaction is static and not modulated in time.
Since we aim to capture the phenomenological features of this setup we consider spectral densities that are simple to treat in the pseudomode approach, while keeping the qualitative features of the setup of Ref.~\cite{Cavaliere2022}.
In particular, we consider a physical scenario in which the two bosonic baths have an antisymmetrized Lorentzian spectral densities of the form
\begin{equation}
J_{\mathrm{AL}}(\gamma, \bar{\omega} ; \omega) = \gamma \left( \frac{1}{\frac{1}{4} \gamma^2 + (\omega - \bar{\omega})^2} - \frac{1}{\frac{1}{4} \gamma^2 + (\omega + \bar{\omega})^2} \right),
\end{equation}
the antisymmetrization ensures that these can indeed represent a physical spectral density describing a continuum of harmonic oscillators.
This class of spectral densities can be approximated by a single thermal pseudomode in certain regimes~\cite{Lemmer2018}, i.e. one can set $n_j=1$ in Eqs.~\eqref{eq:ddt}--\eqref{eq:himj} and \eqref{eq:hmj}--\eqref{eq:njk}, so that
the free dynamics is described by the time-independent Lindblad generator
\begin{equation}
  \mathcal{D} [ \rho ] = - i \bar{\omega} \left[ \widehat{b}^\dag \widehat{b} , \rho \right] + \left( \bar{n} + 1 \right) \left( \widehat{b} \rho \widehat{b}^\dag  - \frac{1}{2} \left\{  \widehat{b}^\dag  \widehat{b} , \rho \right\} \right) + \bar{n} \left( \widehat{b}^\dag \rho \widehat{b}  - \frac{1}{2} \left\{  \widehat{b} \widehat{b}^\dag , \rho \right\} \right),
\end{equation}
with $\bar{n} = ( e^{\bar{\omega}/T} -1 )^{-1}$ where $T$ is the temperature of the surrogate oscillator (again we use units such that $k_B=1$).

\begin{figure}[t]
    \centering
  \includegraphics{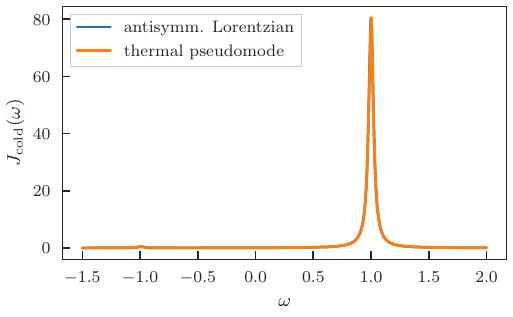}
  \includegraphics{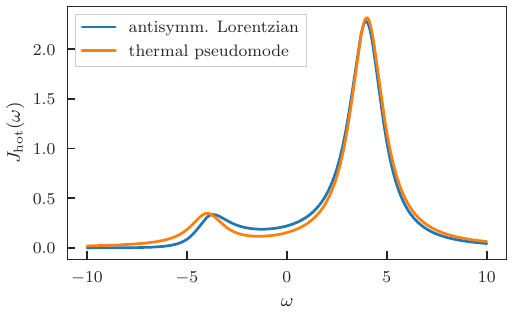}
  \caption{Comparison between the physical antisymmetrized Lorentzian spectral density and the spectral density obtained by an effective description in terms of a single thermal pseudomode.
  Left: cold bath spectral density with a sharp peak at $\bar{\omega}_{\mathrm{cold}}$, parameters $T_{\mathrm{cold}} = 0.2 \bar{\omega}_{\mathrm{cold}}$, $\gamma_{\mathrm{cold}}=0.05 \bar{\omega}_{\mathrm{cold}}$. 
  Right: hot bath with a broader spectral density, parameters $T_{\mathrm{hot}} = 2 \bar{\omega}_{\mathrm{cold}} $, $\gamma_{\mathrm{hot}}=2 \bar{\omega}_{\mathrm{cold}} $ and $\bar{\omega}_{\mathrm{hot}}=4 \bar{\omega}_{\mathrm{cold}}$.
  }
  \label{fig:Cavaliere_spectral_densities}
\end{figure}

We chose the cold bath to have a peak at $\bar{\omega}_{\mathrm{cold}}$, which we keep fixed and use as our unit of time and frequency, a temperature $T_{\mathrm{cold}}= 0.2\bar{\omega}_{\mathrm{cold}}$, and a narrow peak with $\gamma_{\mathrm{cold}} = 0.05 \bar{\omega}_{\mathrm{cold}}$.
The hot bath parameters are chosen as $T_{\mathrm{hot}} = 2 \bar{\omega}_{\mathrm{cold}}$, $\gamma_{\mathrm{hot}} = 2 \bar{\omega}_{\mathrm{cold}}$ and $\bar{\omega}_{\mathrm{hot}} = 4\bar{\omega}_{\mathrm{cold}}$, so that the spectral density in the region around the system frequency does not show a distinct peak.
We show the comparison between the AL thermalized spectral densities, corresponding to a continuum of harmonic oscillators, and the spectral density obtained from a single thermal pseudomode in Fig.~\ref{fig:Cavaliere_spectral_densities}.
Clearly, while the narrow spectral density of the cold bath is well approximated by a thermalized oscillator~\cite{Lemmer2018}, the approximation is not as good for the hot bath.
Nonetheless, we expect the qualitative results to be independent of such details and we proceed to use a single thermal surrogate oscillator as an effective qualitative description of the bath.

The coupling with the cold bath is modulated sinusoidally as
\begin{equation}
  \lambda_{\mathrm{cold}}(t) = g_{\mathrm{cold}} \cos \left(\Omega_d t \right),
\end{equation}
so that the period is $\tau= \frac{2\pi}{\Omega_d}$.
The system Hamiltonian is fixed and time-independent: $H_S =  \omega_0 \frac{ \sigma_z}{2}$.
The resonance condition highlighted in Ref.~\cite{Cavaliere2022} (for weak-coupling) is $\omega_0 = \bar{\omega}_\mathrm{cold} + \Omega_d$.
Concretely, in the numerical simulations we start the dynamics from the excited state of the system $\ket{e}$, i.e. $\sigma_z \ket{e} = \ket{e}$, but this choice does not play a relevant role, since a periodic steady state emerges.

\begin{figure}[t]
  \centering
  \includegraphics{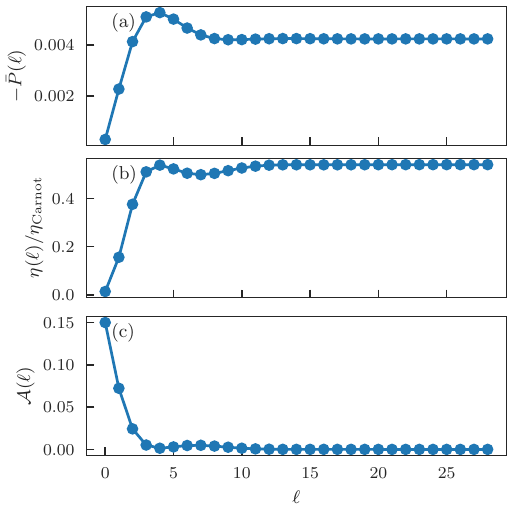}
    \caption{Thermodynamic quantities per cycle as a function of the number of cycles $\ell$. 
    Panel (a): average power per cycle (useful work output corresponds to negative power in our conventions); panel (b): normalized efficiency; panel (c): residual energy. 
    These are obtained for the parameters $g_\mathrm{cold}=0.1$, $g_\mathrm{hot}=0.4$ (relatively strong coupling), $\Omega_d = \bar{\omega}_{\mathrm{cold}} $ and $\omega_0=2 \bar{\omega}_{\mathrm{cold}}$, so that we are at the resonance conditions $\omega_0 = \Omega_d + \bar{\omega}_{\mathrm{cold}}$.
    }
    \label{fig:Cavaliere_work_efficiency}
\end{figure}

To characterize the performance of the thermal machine we compute the total work (equal to the interaction work, since this is the only time-dependent part of the Hamiltonian) per cycle:
\begin{align}
  W_{\mathrm{tot}}(\ell) =  W_{\mathrm{tot}}( \ell \tau , (\ell-1) \tau ) =  W_{I,\mathrm{cold}}(\ell)  = W_{I,\mathrm{cold}}( \ell \tau , (\ell-1) \tau )
\end{align}
and heat exchanged with the two baths per cycle:
\begin{align}
  Q_{\mathrm{hot}}(\ell) =  Q_{\mathrm{hot}}( \ell \tau , (\ell-1) \tau )  \qquad
  Q_{\mathrm{cold}}(\ell)  =  Q_{\mathrm{cold}}( \ell \tau , (\ell-1) \tau ).
\end{align}
The main figures of merit will be the power and efficiency per cycle, defined as
\begin{equation}
  P(\ell) = \frac{W_{\mathrm{tot}}(\ell)}{ \tau} \qquad \eta (\ell) = \frac{W_\mathrm{tot} (\ell)}{Q_\mathrm{hot} (\ell)}.
\end{equation}
Notice that, with our sign conventions, a thermal machine extracting useful work from the temperature difference between the two baths gives negative power values $ P(\ell) $ and negative ${Q_\mathrm{hot} (\ell)}$, i.e. a decrease of the energy of the hot bath, so that the efficiency is positive.
In Fig.~\ref{fig:Cavaliere_work_efficiency} we show these quantities during the dynamics of the thermal machine.
We see that all the quantities settle to the asymptotic value after some initial warm-up cycles. 

As a consistency check, we also compute the variation in system and interaction energy over each period:
\begin{equation}
  \mathcal{A}(\ell) = U(\ell) + \sum_j I_j(\ell), \quad \text{with }   U(\ell) = U( \ell \tau , (\ell-1) \tau ), \text{and }  I_j(\ell) = I_j( \ell \tau , (\ell-1) \tau ). 
\end{equation}
Following Ref.~\cite{Liu2021a} we have also introduced a single quantity $\mathcal{A}(\ell)$, that we dub ``residual energy''.
This is energy that is still ``trapped'' in the system and cannot be converted into work, which would lower the efficiency of the thermal machine.
All these energy differences are expected to vanish when a periodic steady state is reached.
While it has been pointed out that this is not always the case~\cite{Liu2021a}, we show that this happens for our model, as shown for example in Fig.~\ref{fig:Cavaliere_work_efficiency}.

\begin{figure}[t]
  \centering
  \includegraphics{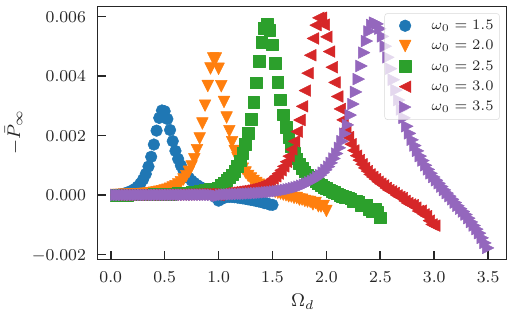}
  \includegraphics{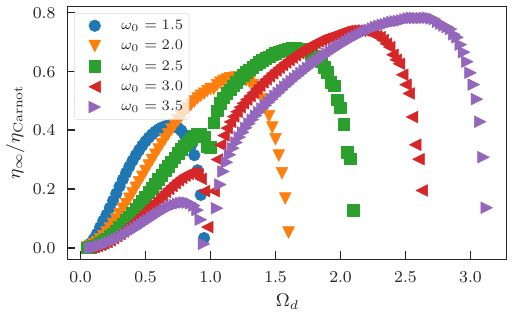}
    \caption{Thermal machine performance as a function of the driving frequency,  for a few different values of the system frequency $\omega_0$.
    Left: average power per cycle (left); right: normalized efficiency.
    Apart from $\Omega_d$ and $\omega_0$, all the other parameters are fixed to the same values as in Fig.~\ref{fig:Cavaliere_work_efficiency}.
    All frequencies are in units of $\bar{\omega}_{\mathrm{cold}}$.
    }
    \label{fig:Cavaliere_asymp_vs_Omegadrive}
\end{figure}

After this brief examination of the dynamics, we move to consider the asymptotic values of the figures of merit $P_\infty$ and $\eta_\infty$ as a function of the system parameters.
Formally these would be the quantities obtained in the limit $\ell \to \infty$, but we found that it is enough to consider $\ell_\mathrm{max} = 20$ periods in the numerical simulations.
In Fig.~\ref{fig:Cavaliere_asymp_vs_Omegadrive} we show the asymptotic power and efficiency for fixed spectral densities of the cold and hot bath, as described above, as a function of the driving frequency and for three different value of TLS frequency.
Since $\bar{\omega}_{\mathrm{cold}}$ remains fixed, in Fig.~\ref{fig:Cavaliere_asymp_vs_Omegadrive} it is possible to observe the system operating as a thermal machine around the resonance condition $\omega_0 = \bar{\omega}_{\mathrm{cold}} + \Omega_d$, which roughly corresponds to the peak in the asymptotic power per cycle.

Overall, these results confirm that it is possible to operate this configuration of baths and sinusoidal driving as a thermal machine also for a TLS, going beyond the harmonic oscillator considered in Ref.~\cite{Cavaliere2022}.
Interestingly, we also observe that the output power and the efficiency improve as the system frequency increases.
This was not explicitly highlighted in Ref.~\cite{Cavaliere2022}, but it may be due to the different spectral densities.
Since we are working with a structured hot bath, differently from Ref.~\cite{Cavaliere2022} where an ohmic hot bath is considered, changing the system frequency is expected to affect the results more strongly.
In particular the interaction with the hot bath should be more pronounced when working close to resonance with the central frequency $\bar{\omega}_{\mathrm{hot}} = 4$ and this may be the cause of the increased performance of the thermal machine.

\section{Conclusions}
\label{sec:conclusions}

In this study, we explored the dynamics and thermodynamic properties of quantum systems coupled to thermal baths, focusing on scenarios beyond the weak coupling regime.
Our first result was to provide a simple derivation of fully general expressions for average thermodynamic quantities in Eq.~\eqref{eq:qexpc}--\eqref{eq:iexpc}, in terms of integrals of expressions involving two-time correlation functions, the bath autocorrelation functions and its time derivative.
From a fundamental point of view, this means that these quantities can be in principle reconstructed from measurements on the system only.
From a practical one, any numerical method able to evaluate such correlation functions can be employed to study quantum thermodynamics nonperturbatively, beyond weak coupling.

Then we specialized our analysis, focusing on the pseudomode approach, demonstrating the possibility to evaluate thermodynamic quantities such as heat, work, and interaction energy through the dynamics of the open system and its pseudomodes.
In particular, we have shown that this approach is effective to study quantum thermodynamics efficiently from a numerical point of view, in accordance with similar results in Refs.~\cite{Lacerda2023a,Lacerda2023,Menczel2024}, since it does not require the evaluation of two-time expectation values of the open system, but only one-time expectation values of the system and pseudomodes.

We have applied this methodology to study a TLS coupled to an ohmic bath, showing that the pseudomode approach can be used to calculate the entropy production.
This was validated by comparing results with known weak-coupling limits, confirming that the pseudomode method effectively handles both strong and weak coupling scenarios.
Finally, we analyzed a simple quantum thermal machine composed of a TLS interacting with two thermal baths at different temperatures.
By modulating the coupling with the cold bath, we demonstrated that work could be extracted efficiently, extending the result of Ref.~\cite{Cavaliere2022} to a finite-dimensional open quantum system.

We also highlight that in the analysis of this application we have empirically observed the emergence of a periodic steady state.
A prospective area for further research is to understand the conditions under which a steady state emerges for the system and pseudomodes, especially in presence of a time-dependent Hamiltonian.
For example, the sufficient conditions found in Ref.~\cite{Menczel2019} do not apply to the standard pseudomode scenario, because the dissipator acts only on the pseudomodes.
On the other hand, embedding techniques have proven effective in studying the steady state of strong-coupling non-Markovian dynamics \cite{Link2024}.

Overall, our findings demonstrate the potential of the pseudomode approach in quantum thermodynamics, offering a robust tool for studying complex quantum systems beyond the traditional weak coupling approximations.
Since the method relies on the solutions of Lindblad dynamics, techniques devised to tackle more complex open systems that have already been studied in other contexts~\cite{Somoza2019} will likely be useful also for this approach.

\section*{Acknowledgments}
We thank Alessandra Colla and Giacomo Guarnieri for fruitful discussions.
The authors acknowledge financial support from Ministero dell'Università e della Ricerca under the ``PON Ricerca e Innovazione 2014-2020'' project EEQU.

\appendix

%%%%%%%%%%%%
\section{Derivation of thermodynamic quantities in terms of two-time correlation functions}\label{app:derivation_2time}

In this Appendix, we derive the expression of heat, work, and interaction energies of Eqs.~\eqref{eq:qexpc}--\eqref{eq:iexpc} that depend only on the bath correlation function and two-time expectation values of system operators, relying on the Heisenberg picture evolution of the global unitary dynamics, along with a unitary transformation on the environmental modes.

\subsection{Thermofield mapping}
The first step is to use a Bogoliubov transformation, known as thermofield mapping, on the environmental modes to map a bath initially in a thermal state at finite temperature into a zero-temperature bath~\cite{Diosi1998,Yu2004,deVega2015b,Tamascelli2019}.
This mapping will be applied independently to each bath, since they are initially uncorrelated and there are no interaction terms in the Hamiltonian.
Hence, we consider a system and a single bath with the Hamiltonian specified in Eq.~\eqref{eq:mainh} (for a single value of $j$) and the initial global state in Eq.~\eqref{eq:rhoSB0}, which we rewrite here for convenience: 
\begin{equation}\label{eq:rho1beta}
\rho^{(1)}_{SB}(0) = \rho_S(0) \otimes \rho_B^{\beta};
\end{equation}
in this section we denote with the label $(1)$ the states corresponding to the original configuration,
to keep track of the difference with the configuration resulting from the thermofield mapping.
The state at time $t$ can be formally written as
\begin{equation}
\rho^{(1)}_{SB}(t) = \widehat{U}^{(1)}_{SB}(t) \rho^{(1)}_{SB}(0)\widehat{U}^{(1) \dag}_{SB}(t); \qquad 
\widehat{U}^{(1)}_{SB}(t) = T e^{-i \int_0^t d s \widehat{H}^{(1)}_{SB}(s)}. 
\end{equation}
Consider now a second bosonic bath $B'$ such that the joint system $S$--$B$--$B'$ is ruled by the Hamiltonian
\begin{equation}
\widehat{H}^{(2)}_{SBB'}(t) = \widehat{H}^{(1)}_{SB}(t)\otimes \mathbbm{1}_{B'}+\mathbbm{1}_{SB}\otimes \widehat{H}_{B'}
\end{equation}
with 
\begin{equation}
\widehat{H}_{B'} = -\sum_k \omega_k \widehat{c}^\dag_k \widehat{c}_k,
\end{equation}
with $\widehat{c}_k$ and $\widehat{c}^\dag_k$ the annihilation and creation operators associated with the $B'$ mode with frequency $\omega_k$ (and the set of possible frequencies is the same as the one of the modes $\widehat{b}_k, \widehat{b}^\dag_k$).
The corresponding global unitary operator has indeed the factorized form
\begin{equation}\label{eq:facu}
\widehat{U}^{(2)}_{SBB'}(t) = \widehat{U}^{(1)}_{SB}(t)\otimes \widehat{U}_{B'}(t); \qquad \widehat{U}_{B'}(t) = e^{- i \widehat{H}_{B'}t}
\end{equation}
and thus we have
\begin{equation}\label{eq:eqt}
\Tr_{SBB'}\left\{\widehat{O}_{SB}(t)\rho^{(2)}_{SBB'}(t)\right\} =
\Tr_{SBB'}\left\{\widehat{O}_{SB}(t)\widehat{U}^{(2)}_{SBB'}(t)\rho^{(2)}_{SBB'}(0)\widehat{U}^{(2)\dag}_{SBB'}(t)\right\} = 
\Tr_{SB}\left\{\widehat{O}_{SB}(t)\rho^{(1)}_{SB}(t)\right\}
\end{equation}
for any initial state $\rho^{(2)}_{SBB'}(0)$ such that
\begin{equation}\label{eq:eqini}
\Tr_{B'}\left\{\rho^{(2)}_{SBB'}(0)\right\} = \rho^{(1)}_{SB}(0),
\end{equation}
and for any, possibly time-dependent, operator $\widehat{O}_{SB}(t)$ on the $S$--$B$ subsystem. 
As a consequence, any expectation value on the original $S$--$B$ configuration can be equivalently calculated in the auxiliary $S$--$B$--$B'$ configuration, provided that Eq.~\eqref{eq:eqini} holds. 
As we will now show, this can be particularly convenient after a suitable unitary transformation on $S$--$B$--$B'$.

Defining the new modes  $\widehat{d}_k,\widehat{e}_k$ via the Bogoliubov transformation
\begin{eqnarray}
 \widehat{b}_k&=&\sqrt{n_k+1}\widehat{d}_k+\sqrt{n_k}\widehat{e}^\dag_k \notag\\
\widehat{c}_k&=&\sqrt{n_k+1}\widehat{e}_k+\sqrt{n_k}\widehat{d}^\dag_k, \label{eq:bog}
\end{eqnarray}
with
\begin{equation}
n_k = \frac{1}{e^{\beta \omega_k}-1},
\end{equation}
the global Hamiltonian $\widehat{H}^{(2)}_{SBB'}(t)$ can be equivalently written as
\begin{eqnarray}
\widehat{H}^{(2)}_{SBB'}(t) &=&  \widehat{H}_S(t) 
+ \sum_k \omega_k \widehat{d}^\dag_k \widehat{d}_k
- \sum_k \omega_k \widehat{e}^\dag_k \widehat{e}_k
+ \lambda(t)\widehat{A} \otimes \left(
\sum_k g_k \left(\sqrt{n_k+1}
(\widehat{d}_k+\widehat{d}^\dag_k)+\sqrt{n_k}
(\widehat{e}_k+\widehat{e}^\dag_k)\right)\right).\label{eq:thermoh}
\end{eqnarray}
Now, consider the vacuum state of $\widehat{d}_k$ and $\widehat{e}_k$, 
$\widehat{d}_k\ket{\Omega} = \widehat{e}_k\ket{\Omega} = 0$,
as the initial state of $B$--$B'$,
so that
\begin{equation}\label{eq:invac}
\rho^{(2)}_{SBB'}(0) = \rho_{S}(0) \otimes \ket{\Omega}\bra{\Omega}.
\end{equation}
Importantly, $\ket{\Omega}\bra{\Omega}$ is a purification of $\rho_B^{\beta}$, i.e.,~\cite{deVega2015b}
\begin{equation}
\rho_B^{\beta} = \Tr_{B'}\left\{\ket{\Omega}\bra{\Omega}\right\},
\end{equation} 
so that Eqs.~\eqref{eq:eqt} and~\eqref{eq:eqini} tell us that the expectation value of any $S$--$B$ operator can be equivalently evaluated by replacing the bath with a double set of modes that are initially in the vacuum, but whose coupling with the system is weighted by two factors depending on the temperature of the original configuration.
A further simplification can be made by introducing a unique family of creation and annihilation operators; let $I$ be the set of values over which $k$ varies in Eq.~\eqref{eq:mainh}, and thus in Eq.~\eqref{eq:thermoh}, and let us formally double it in $I_1$ and $I_2$ (i.e., $I_1=I_2=I$) and define the set $J =I_1\cup I_2$.
Then, introducing the annihilation and creation operators $\widehat{f}_k, \widehat{f}^\dag_k$, with $k$ taking values in $J$, via
\begin{equation}
\widehat{f}_k = \begin{cases} \widehat{d}_k \quad \mbox{if} \quad k\in I_1\\
\widehat{e}_k \quad \mbox{if} \quad k\in I_2, \label{eq:bogf}
\end{cases}
\end{equation}
along with
\begin{equation}
\omega'_k = \begin{cases} \omega_k \quad \mbox{if} \quad k\in I_1\\
-\omega_k \quad \mbox{if} \quad k\in I_2,
\end{cases}
\end{equation}
and the couplings
\begin{equation}\label{eq:thermg}
g'_k = \begin{cases} g_k \sqrt{n_k+1}  \quad \mbox{if} \quad k\in I_1\\
g_k \sqrt{n_k}  \quad \mbox{if} \quad k\in I_2,
\end{cases}
\end{equation}
the Hamiltonian in Eq.~\eqref{eq:thermoh} can be written as
\begin{eqnarray}
\widehat{H}^{(2)}_{SBB'}(t) &=&  \widehat{H}_S(t) 
+ \sum_{k\in J} \omega'_k \widehat{f}^\dag_k \widehat{f}_k
+ \lambda(t)\widehat{A} \otimes
\sum_{k\in J} g'_k \left(\widehat{f}_k+\widehat{f}^\dag_k\right), \label{eq:thermohfin}
\end{eqnarray}
which takes the same form as the Hamiltonian of the initial configuration, apart from the different set of values taken by $k$.
Crucially, the coupling coefficients $g'_k$ now encode the dependence on the bath temperature due to Eq.~\eqref{eq:thermg}.
In the following we will refer to the operators $\{ \widehat{f}_k \}_k$ as the \emph{thermalized} annihilation operators, or thermalized modes.

Finally, given the free environmental Hamiltonian $\widehat{H}_{BB'} =  \sum_{k\in J} \omega'_k \widehat{f}^\dag_k \widehat{f}_k$ and the environmental interaction operator $\widehat{X}' = \sum_{k\in J} g'_k \left(\widehat{f}_k+\widehat{f}^\dag_k\right)$, the autocorrelation of the latter is defined as (cf. Eq.~\eqref{eq:cj})
\begin{equation}\label{eq:scorrp}
 C'(t) = \langle X'(t) X'(0) \rangle  = \bra{\Omega}e^{i \widehat{H}_{BB'} t}\widehat{X}' e^{-i \widehat{H}_{BB'} t} \widehat{X}'\ket{\Omega}.
\end{equation}
Introducing the spectral density for the zero-temperature bath, with the couplings as in Eq.~\eqref{eq:thermg}
\begin{equation}\label{eq:jprime}
J'(\omega) =  \sum_{k\in J} g'^2_{k} \delta(\omega-\omega'_{k}),
\end{equation}
one immediately sees that (compare with Eq.~\eqref{eq:autocorr_thermal_bath})
\begin{equation}\label{eq:cprimet}
    C'(t) = \int_{-\infty}^{\infty} d \omega J'(\omega) e^{-i \omega t}.
\end{equation}
But then the identity
\begin{equation}
\int_{-\infty}^{\infty} d \omega J'(\omega) e^{-i \omega t} = \int_0^\infty d \omega J(\omega) \left(  \cos(\omega t) \coth \left(\frac{\beta\omega}{2} \right)  - i \sin ( \omega t )\right) 
\end{equation}
allows us to conclude that
\begin{equation}\label{eq:samecorr}
    C(t) = C'(t),
\end{equation}
i.e., one has the same autocorrelation functions for the environmental interaction operators, respectively, in the configuration with an initial thermal state at finite temperature, and in the configuration with a zero-temperature environment---each one with respect to their free environmental dynamics. 
This is the property at the basis of the definition of the so-called thermalized TEDOPA~\cite{Tamascelli2019} and $J'(\omega)$ corresponds to the thermalized spectral density $J_\beta(\omega)$ introduced in Eq.~\eqref{eq:autocorr_thermal_bath}.

\subsection{Connection with open-system two-time correlation functions}\label{sec:cwo}

We now proceed by recalling the definitions of heat and work for the original configuration.
Using
\begin{equation}
  \frac{d \rho_{SB}(t)}{d t} = - i \left[ \widehat{H}_{tot}(t), \rho_{SB}(t)\right]
\end{equation}
and the cyclicity of the trace, the heat in Eq.~\eqref{eq:uiq} can be written as (with $t_i=0$)
\begin{equation}\label{eq:qjtau2}
Q_j(t_f) = \int_0^{t_f} d t \Tr\left\{\widehat{H}_{B,j} \frac{d}{d t}\rho_{SB}(t)\right\} = i \int_0^{t_f} d t \Tr\left\{\left[\widehat{H}_{tot}(t), \widehat{H}_{B,j}\right] \rho_{SB}(t)\right\} = i \int_0^{t_f} d t \Tr\left\{\left[\widehat{H}_{I,j}(t), \widehat{H}_{B,j}\right] \rho_{SB}(t)\right\},
\end{equation}
so that Eqs.~\eqref{eq:mainh} and~\eqref{eq:can} directly lead us to
\begin{equation}\label{eq:preq}
Q_j(t_f) = i \sum_k g_{j,k}\omega_{j,k} \int_0^{t_f} d t \lambda_j(t)
\Tr\left\{\left(\widehat{A}_{S,j} \otimes 
(\widehat{b}_{j,k} - \widehat{b}^\dag_{j,k})\right) \rho_{SB}(t)\right\}.
\end{equation}

Now, the key point is that the expectation value of the system-environment operator  $\widehat{A}_{S,j} \otimes (\widehat{b}_{j,k} - \widehat{b}^\dag_{j,k})$ can be actually turned into a two-time correlation function of the open-system operator $\widehat{A}_{S,j}$.
To show this, we first exploit the thermofield mapping.
Using Eqs.~\eqref{eq:facu}--\eqref{eq:bog} and \eqref{eq:bogf}--\eqref{eq:thermg}, the heat exchanged up to a time $t_f$ with the $j$-th bath $Q_j(t_f)$ in Eq.~\eqref{eq:preq} can be expressed with respect to the thermalized creation and annihilation operators as
\begin{equation}\label{eq:preqth}
Q_j(t_f) = i \sum_{k\in J} g'_{j,k}\omega'_{j,k} \int_0^{t_f} d t \lambda_j(t)
\Tr\left\{\left(\widehat{A}_{S,j} \otimes 
(\widehat{f}_{j,k} - \widehat{f}^\dag_{j,k})\right) \rho_{SBB'}^{(2)}(t)\right\},
\end{equation}
where $\rho_{SBB'}^{(2)}(t)$ denotes the global state of the system made of $S$, $B_j$ and $B'_j$, for $j=1,\ldots, N_B$ obtained from the initial product state
\begin{equation}\label{eq:inkk}
 \rho_{SBB'}^{(2)}(0)=\rho_S(0)\bigotimes_{j=1}^{N_B}\ket{\Omega_j}\bra{\Omega_j},
\end{equation}
with $\ket{\Omega_j}$ the vacuum state of the annihilation operators $\widehat{d}_{j,k}, \widehat{e}_{j,k}$ of the bosonic systems $B_j$ and $B'_j$, see Eq.~\eqref{eq:invac}.
Moving to the Heisenberg picture, see Eq.~\eqref{eq:eut}, we have to evaluate
\begin{equation}\label{eq:newb}
\Tr\left\{\left(\widehat{A}_{S,j}(t) 
(\widehat{f}_{j,k}(t) - \widehat{f}^\dag_{j,k}(t))\right) \rho^{(2)}_{SBB'}(0)\right\}
= 2 i \Im\left\{\Tr\left\{\widehat{A}_{S,j}(t) 
\widehat{f}_{j,k}(t) \rho^{(2)}_{SBB'}(0)\right\}\right\},
\end{equation}
where $\Im$ denotes the imaginary part and the identity follows from the fact that (unitary evolutions preserve observables commutativity)
\begin{equation}
\left[\widehat{A}_{S,j}(t), \widehat{f}^\dag_{j,k}(t)\right]=0.
\end{equation}
Then, we use the solution of the Heisenberg equation fixed by Eq.~\eqref{eq:thermohfin} (straight-forwardly generalized to the case of $N_B$ independent baths) for $\widehat{f}_{j,k}(t)$, 
\begin{equation}\label{eq:bhei}
\widehat{f}_{j,k}(t) = 
e^{-i \omega'_{j,k}t} \widehat{f}_{j,k}- i g'_{j,k} \int_0^t ds \lambda_j(s) e^{-i \omega'_{j,k}(t-s)}
\widehat{A}_{S,j}(s),
\end{equation}
and we exploit once more the properties of the thermalized modes, for which the baths are initially in the vacuum state of the $\widehat{f}_{j,k}$ fields, so that the contribution in Eq.~\eqref{eq:preqth} due to the first term in Eq.~\eqref{eq:bhei} vanishes.
Hence, introducing the two-time correlation function of the open system observable $O_S(t)$ defined as
\begin{equation}\label{eq:twotime}
\left\langle O_S(t)O_S(s)\right\rangle = \Tr\left\{\widehat{O}_S(t)\widehat{O}_S(s)\rho^{(2)}_{SBB'}(0)\right\}, 
\end{equation}
we get
\begin{eqnarray}
Q_j(t_f) &=& 
2 \sum_{k\in J} g'^2_{j,k}\omega'_{j,k} \int_0^{t_f} d t \lambda_j(t)
\int_0^t ds  \lambda_j(s) \Re\left\{e^{-i \omega'_{j,k}(t-s)}
\left\langle A_{S,j}(t) A_{S,j}(s)\right\rangle\right\}. \label{eq:qexp}
\end{eqnarray}
Introducing the spectral density defined in Eq.~\eqref{eq:jprime} and taking the continuum limit, we get
\begin{eqnarray}
Q_j(t_f) &=& 2\int^{\infty}_{-\infty} d \omega J'_{j}(\omega) \omega \int_0^{t_f} d t \lambda_j(t)
 \int_0^t ds \lambda_j(s)\Re\left\{e^{-i \omega(t-s)}
\left\langle A_{S,j}(t) A_{S,j}(s)\right\rangle\right\}\notag
\end{eqnarray}
and then, using Eq.~\eqref{eq:cprimet},
\begin{eqnarray}
Q_j(t_f) &=&- 2 \int_0^{t_f} d t \lambda_j(t)
\int_0^t ds \lambda_j(s) \Im\left\{\dot{C}_j(t-s)
\left\langle A_{S,j}(t) A_{S,j}(s)\right\rangle\right\} \notag.\label{eq:qexpcapp}
\end{eqnarray}

We note that because of Eq.~\eqref{eq:samecorr}---and since the expectation value of the environmental interaction operators on the initial states vanish---the equivalence theorem recalled in Sec.~\ref{sec:tpa} guarantees that the autocorrelation function in Eq.~\eqref{eq:twotime} for $\widehat{O}_S = \widehat{A}_{S,j}$ coincides with the one in Eq.~\eqref{eq:twotimemain}, which then concludes our proof of Eq.~\eqref{eq:qexpc}.

Analogously, starting from Eq.~\eqref{eq:wwi} (cf. Eq.~\eqref{eq:preq}) and using the thermalized modes, along with 
\begin{equation}\label{eq:newb2}
\Tr\left\{\left(\widehat{A}_{S,j}(t) 
(\widehat{f}_{j,k}(t) + \widehat{f}^\dag_{j,k}(t))\right) \rho^{(2)}_{SBB'}(0)\right\}
= 2 \Re\left\{\Tr\left\{\widehat{A}_{S,j}(t) 
\widehat{f}_{j,k}(t) \rho^{(2)}_{SBB'}(0)\right\}\right\},
\end{equation}
and Eq.~\eqref{eq:bhei}, we can express the work due to the time-dependent interaction Hamiltonian as
\begin{equation}\label{eq:wexp}
W_I(t_f) =2 \sum_{k\in J} g'^2_{j,k} \int_0^{t_f} d t \frac{d\lambda_j(t)}{dt}
\int_0^t ds \lambda_j(s) \Im\left\{e^{-i \omega'_{j,k}(t-s)}
\left\langle A_{S,j}(t) A_{S,j}(s)\right\rangle\right\},
\end{equation}
so that, following similar steps as in the previous derivation, one arrives at Eq.~\eqref{eq:wexpc}.
Finally, starting from Eq.~\eqref{eq:ui} and using
  \begin{equation}
    \label{eq:HintExp}
    \Tr[ \widehat{H}_{I,j} (t) \rho_{SB}(t) ] =  \lambda_j(t) \Tr\left\{ \left[ \widehat{A}_{S,j}\otimes\sum_k g_{j, k}\left(\widehat{b}_{j,k}+\widehat{b}^\dag_{j,k}\right) \right] \rho_{SB}(t) \right\},
  \end{equation}
one can similarly arrive at an expression for the interaction energy
  \begin{equation}
      \Tr[ \widehat{H}_{I,j} (t) \rho_{SB}(t) ] = 2 \lambda_j(t) \int_0^t ds \lambda_j(s) \Im\left\{ C_j(t-s)  \left\langle A_{S,j}(t) A_{S,j}(s) \right\rangle\right\},
  \end{equation}
and then to Eq.~\eqref{eq:iexpc}, since the initial expectation value of $\widehat{H}_I$ is zero, as the state is thermal.

\section{Proof of the equivalence for a generic product of environmental interaction operators}\label{app:pef}

Consider two unitary configurations with Hamiltonians as in Eq.~\eqref{eq:mainh} and with the same open system and number of baths, i.e.,
\begin{eqnarray}
  \widehat{H}_{SB}(t) &=& \widehat{H}_S(t)
  +\sum^{N_B}_{j=1} \widehat{H}_{B,j}
  +\sum^{N_B}_{j=1} \widehat{H}_{I,j}(t); \nonumber\\
    \widehat{H}'_{SB'}(t) &=& \widehat{H}_S(t)
  +\sum^{N_{B'}}_{j=1} \widehat{H}'_{B',j}
  +\sum^{N_{B'}}_{j=1} \widehat{H}'_{I,j}(t)\label{eq:b1}
\end{eqnarray}
with
\begin{eqnarray}
  \widehat{H}_{I,j}(t)&=&\lambda_j(t) \widehat{A}_{S,j}\otimes \widehat{X}_j,
  \qquad \widehat{X}_j = \sum_k g_{j, k}\left(\widehat{b}_{j,k}+\widehat{b}^\dag_{j,k}\right);\nonumber\\  
  \widehat{H}'_{I,j}(t)&=&\lambda_j(t) \widehat{A}_{S,j}\otimes \widehat{X}'_j, 
  \qquad \widehat{X}'_j=\sum_k g'_{j, k}\left(\widehat{b}'_{j,k}+\widehat{b}'^\dag_{j,k}\right),\nonumber
\end{eqnarray}
where $\widehat{b}_{j,k}$ and $\widehat{b}^{\dag}_{j,k}$ are the bosonic creation and annihilation operators of the $j$-th bath in the environment $B$, while $\widehat{b}'_{j,k}$ and $\widehat{b}'^{\dag}_{j,k}$ are referred to the environment $B'$.
The environmental free Hamiltonians $\widehat{H}_{B,j}$ and $\widehat{H}'_{B',j}$ are at most quadratic in the corresponding creation and annihilation operators.
Moreover, consider the initial states $\rho_{SB}(0)=\rho_{S}(0)\otimes\rho_{B}(0)$ and $\rho_{SB'}(0)=\rho_{S}(0)\otimes\rho_{B'}(0)$, where $\rho_{B}(0)$ and $\rho_{B'}(0)$ are thermal states for their free Hamiltonians, so that they are stationary with respect to their corresponding free dynamics
and the expectation values of the environmental interaction operators are zero (the analysis can be easily generalized to deal with generic initial Gaussian states).

Given the autocorrelation functions of the environmental interaction operators,
\begin{eqnarray}
    C_j(t) &=& \Tr_B\left[e^{i \widehat{H}_B t}\widehat{X}_j e^{-i \widehat{H}_B t}\widehat{X}_j  \rho_B(0)\right] \nonumber\\
    C'_{j}(t) &=& \Tr_{B'}\left[e^{i \widehat{H}'_{B'} t}\widehat{X}'_{j} e^{-i \widehat{H}'_{B'} t}\widehat{X}'_{j} 
    \rho_{B'}(0)\right],\label{eq:impl1}
\end{eqnarray}
if they coincide, then the expectation values of system-environment operators involving a generic product of the environmental interaction operators coincide as well, i.e.,
\begin{equation}
 C_j(t) = C'_{j}(t) \quad \forall j,\, t \quad \Longrightarrow \quad 
\Tr_{SB}\left[\widehat{O}_S\otimes \widehat{X}_{j_1}\ldots\widehat{X}_{j_k} \rho_{SB}(t) \right]
=\Tr_{SB'}\left[\widehat{O}_S\otimes \widehat{X}'_{j_1}\ldots\widehat{X}'_{j_k} \rho_{SB'}(t) \right],\label{eq:impl2}
\end{equation}
where $\widehat{O}_S$ is a generic open-system operator.  
This implication directly follows from taking, for example, a Dyson expansion in the expectation value involving the power of the environmental interaction operator: given a Hamiltonian as in Eq.~\eqref{eq:b1} and working in the interaction picture, we have
\begin{eqnarray}
  &&   \Tr_{SB}\left[\widehat{O}_S\otimes \widehat{X}_{j_1}\ldots\widehat{X}_{j_k} \rho_{SB}(t) \right] \\
  \hspace{-4cm}    &&= \Tr_{SB}\left[\widehat{O}_S(t)\otimes \widehat{X}_{j_1}(t)\ldots\widehat{X}_{j_k}(t) 
     T_{\leftarrow}e^{-i \sum_{j=1}^{N_B} \int_0^t  \widehat{A}_{S,j}(s) \otimes \widehat{X}_j(s)} 
     \rho_{S}(0)\otimes\rho_{B}(0) T_{\rightarrow}e^{i \sum_{j=1}^{N_B} \int_0^t \widehat{A}_{S,j}(s) \otimes \widehat{X}_j(s) }\right] \nonumber
\end{eqnarray}
with, indeed,
\begin{eqnarray}
    \widehat{O}_S(t) =  T_{\rightarrow}e^{i \int_0^t  \widehat{H}_{S}(s)} \widehat{O}_S  T_{\leftarrow}e^{-i \int_0^t  \widehat{H}_{S}(s)}
\end{eqnarray}
(and analogously for $\widehat{A}_{S,j}(t)$), while 
\begin{eqnarray}
    \widehat{X}_j(t) = e^{i \widehat{H}_B t} \widehat{X}_j  e^{-i \widehat{H}_B t};
\end{eqnarray}
expanding the time-ordering operator, we get
\begin{eqnarray}
&&  \Tr_{SB}\left[\widehat{O}_S\otimes \widehat{X}_{j_1}\ldots\widehat{X}_{j_k} \rho_{SB}(t) \right] = \sum_{k',k''=0}^{\infty}(-i)^{k'+k''}\int_0^t d t'_{k'} \int_0^{t'_{k'}} dt'_{k'-1}\ldots \int_0^{t'_2} dt'_{1}
  \int_0^t d t''_{k''} \int_0^{t''_{k''}} d t''_{k''-1}\ldots \int_0^{t''_2} dt''_{1}
   \nonumber\\
&&  \times \sum_{j'_1,\ldots,j'_{k'}=1}^{N_B}\sum_{j''_1,\ldots,j''_{k''}=1}^{N_B} \Tr_{S}\left[\widehat{O}_S\widehat{A}_{S,j'_{k'}}(t'_{k'})\ldots \widehat{A}_{S,j'_1}(t'_1)\rho_{S}(0)\widehat{A}_{S,j''_1}(t''_1)\ldots \widehat{A}_{S,j''_{k''}}(t''_{k'}) \right]\nonumber\\
&&  
\times\Tr_{B}\left[\widehat{X}_{j_1}\ldots\widehat{X}_{j_k}
\widehat{X}_{j'_{k'}}(t'_{k'})\ldots \widehat{X}_{j'_1}(t'_1)\rho_{B}(0)\widehat{X}_{j''_1}(t''_1)\ldots \widehat{X}_{j''_{k''}}(t''_{k'})
\right],
\end{eqnarray}
so that due to the Gaussianity of the bath the latter factor can be univocally written in terms of the autocorrelation functions $C_j(t)$, and then Eq.~\eqref{eq:impl2} follows.
Note that this is the case only because the environmental operator is a product of interaction operators
$\widehat{X}_{j_1}\ldots\widehat{X}_{j_k}$, while expectation values of different operators would not be directly relatable to $C_j(t)$.

To prove the validity of Eq.~\eqref{eq:eqpre}---i.e., that the expectation values of system operators times the composition of environmental interaction operators can be evaluated via the network of pseudomodes defined by Eqs.~\eqref{eq:ddt}--\eqref{eq:himj} provided that Eq.~\eqref{eq:eqq} holds---it is enough to apply Lemma 1 and Lemma 2 of Ref.~\cite{Tamascelli2017}. 
Directly adapting them to the framework at hand here, Lemma 1 can be formulated stating that there is a Hamiltonian $\widehat{H}'_{SM\widetilde{E}}(t)$ as in Eq.~\eqref{eq:b1}, with the identification $B' \mapsto M\text{--}\widetilde{E}$,  on the system, the pseudomodes and some further degrees of freedom $\widetilde{E}$ such that the system-pseudomode evolution can be obtained as the exact reduced dynamics from the unitary evolution on $S\text{--}M\text{--}\widetilde{E}$, and then in particular
\begin{equation}
 \Tr_{SM\widetilde{E}}\left[\widehat{O}_S\otimes \widehat{X}'_{j_1}\ldots\widehat{X}'_{j_k} \rho_{SM\widetilde{E}}(t) \right]
   = \Tr_{SM}\left[\widehat{O}_S\otimes \widehat{X}_{M,j_1}\ldots\widehat{X}_{M,j_k} \rho_{SM}(t) \right].
   \label{eq:impl3}
\end{equation}
Furthermore, denoting as $C'_j(t)$ the autocorrelation functions of the configuration $S\text{--}M\text{--}\widetilde{E}$, Lemma 2 in Ref.~\cite{Tamascelli2017} tells us that they coincide with the pseudomode autocorrelation functions in Eq.~\eqref{eq:cmjj}, i.e.,
\begin{equation}\label{eq:cpcm}
C'_j(t)=C_{M,j}(t);
\end{equation}
thus, Eq.~\eqref{eq:impl2}, with the identification $B' \mapsto M\text{--}\widetilde{E}$, and Eqs.~\eqref{eq:impl3} and \eqref{eq:cpcm} lead us to
\begin{equation}
 C_j(t) = C_{M,j}(t) \quad \forall j,\,\, t\geq0 \quad \Longrightarrow \quad 
 \Tr_{SB}\left[\widehat{O}_S\otimes \widehat{X}_{j_1}\ldots\widehat{X}_{j_k} \rho_{SB}(t) \right]
   = \Tr_{SM}\left[\widehat{O}_S\otimes \widehat{X}_{M,j_1}\ldots\widehat{X}_{M,j_k} \rho_{SM}(t) \right], 
   \label{eq:impl4}
\end{equation}
which concludes the proof.

\bibliography{thermo_bib.bib}

\end{document}